\pdfoutput=1

\documentclass[preprint,authoryear,12pt,times]{elsarticle}

\usepackage{amssymb}
\usepackage{amsmath}
\usepackage{comment}
\usepackage{url}
\usepackage[]{subcaption}
\usepackage{bm}

\def\clap#1{\hbox to 0pt{\hss#1\hss}}

\def\mathrlap{\mathpalette\mathrlapinternal}

\def\mathrlapinternal#1#2{%
	\rlap{$\mathsurround=0pt#1{#2}$}}

\journal{J. Comput. Phys.}

\begin{document}

\begin{frontmatter}



\title{
	Particle mesh multipole method:
   	An efficient solver for 
	gravitational/electrostatic forces
	based on multipole method and fast convolution over a uniform mesh}


\author{Keigo Nitadori}

\ead{keigo@riken.jp}

\address{
	Co-Design Team, Exascale Computing Project, \\
	RIKEN Advanced Institute for Computational Science, \\
	7-1-26, Minatojima-minami-machi, Kobe, Japan
}

\begin{abstract}
We propose an efficient algorithm for the evaluation of the potential and
its gradient of gravitational/electrostatic $N$-body systems,
which we call particle mesh multipole method (PMMM or PM$^3$).
PMMM can be understood both as an extension of the particle
mesh (PM) method and as an optimization of the fast multipole
method (FMM).
In the former viewpoint, 
the scalar density and potential 
held by a grid point 
are extended to multipole moments and local
expansions in $(p+1)^2$ real numbers, where $p$ is the order
of expansion.
In the latter viewpoint,
a hierarchical octree structure which brings
its $\mathcal O(N)$ nature, is replaced with a uniform mesh
structure, and we exploit the convolution theorem with fast
Fourier transform (FFT) to speed up the calculations.
Hence, independent $(p+1)^2$ FFTs
with the size  equal to the number of grid points are performed.

The fundamental idea is common to PPPM/MPE by
\cite{Shimada:1993:ECC}
and FFTM by
\cite{Ong:2003:FAT}.
PMMM differs from them in supporting both the open and periodic 
boundary conditions,
and employing an irreducible form  where both the multipole moments and 
local expansions are expressed in $(p+1)^2$ real numbers
and the transformation matrices in $(2p+1)^2$ real numbers.

The computational complexity is the larger of
$\mathcal O(p^2 N)$ and $\mathcal O(N \log (N/p^2))$,
and the memory 
demand is
$\mathcal O(N)$ when the
number of grid points is 
$\propto N/p^2$.
\end{abstract}


\end{frontmatter}


\section{Introduction}
Particle mesh (PM) methods
[including the particle--particle particl--mesh \allowbreak 
(PPPM or P$^3$M) method \citep{Hockney:1988:CSU},
and the particle mesh Ewald (PME) method \citep{Darden:1993:PME}],
and the fast multipole method (FMM) by 
\cite{Greengard:1987:FAP, Greengard:1988:REPb}
have been developed 
to speed up the calculation of 
long range forces in particle simulations.
An efficient unification of these two methods was accomplished
by \cite*{Shimada:1993:ECC} as a multipole expansion version of
PPPM (PPPM/MPE) to improve the accuracy 
in 
periodic boundary conditions.
Later, they demonstrated the performance advantage over 
PPPM/MPE to the FMM in low-order cases \citep{Shimada:1994:PFM}.
Their multipole formulation was based on direct
Cartesian gradients where the number of multipole terms 
scales as $p^3$ for the order of expansion $p$.
A version with an  irreducible form based on spherical harmonics
where the number of terms is $(p+1)^2$ was introduced by
\cite{Ong:2003:FAT} for open boundary conditions, 
as fast Fourier transform on multipoles (FFTM).

These hybrid schemes have several practical advantages over
both the PPPM and FMM.
Compared to PPPM, fairly high accuracy can be easily
achieved. 
Since a grid point holds 
the information of charge distribution and the potential field
as multipole moments and local expansions,
a particle
needs to interact only with the nearest grid point. 
In the PM methods,
interaction with the nearest $(p+1)^3$ grid points is necessary.
Moreover, relatively coarse mesh can be used and multiple FFTs
with small size can be performed independently
instead of one big FFT. 
This is suitable for cache based and distributed parallel computers of today.
For the particle--particle interaction part, a simple Newtonian
force can be used. 
A cutoff function, which requires
an expensive table look-up or mathematical functions,
is not necessary.
The advantage over FMM is the arithmetic operation cost.
In the case of the periodic boundary, the hybrid scheme
requires only one multipole-to-local (M2L) transformation
per cell 
(a factor 8 overhead exists, however, for open boundary systems).
In the octree-based FMM, each cell
requires 189 transformations when the minimum separation between cells
is set to one cell size (27 cells cutoff), and it increases to 875 when
the minimum separation is set to two (125 cells cutoff) for better
accuracy.
In the hybrid scheme, the number of M2L transformations 
remains constant irrespective of the cell separation criterion,
and this relaxes the motivation to use very high order of expansion $p$.
Thus, the M2L transformation is usually not the bottleneck.
Another practical advantage is that any  products of three integers
can be used for the number of cells, while it is usually limited to
powers of 8 in FMM.

Despite the above benefits, 
the hybrid approach of PM and FMM 
does not seem to be widely used in the
particle simulation field, 
though similar approaches are used in other fields
\citep{,CGFMMFFT2007, Hesford:2010:FMM}.
One reason might be that the formulation is complicated,
and another is that algorithms in previous studies are either
sub-optimal (not irreducible) or limited in applicability
(e.g. only for open boundary).
In this paper, we present an optimal and general scheme based
on this approach, which we named
particle mesh multipole method (PMMM)
\footnote{
	A scheme named pseudoparticle multipole method (PPMM or P$^2$M$^2$) exists as well
	\citep{Makino:1999:YAF}.
}.
In PMMM, a simple solid harmonics notation \citep{Wang:1996:EFM}
is employed for the multipole transformations, 
and both multipole moments and local expansions are expressed
in $(p+1)^2$ real numbers. 
A matrix for a 
$\mathbb R^{(p+1)^2} \to \mathbb R^{(p+1)^2}$
transformation has effectively 
$(2p+1)^2$ real numbers, not $(p+1)^4$. 
In this way, the computational cost and memory
requirements are minimized.
As far as we know, ours is the first implementation that supports
a periodic boundary condition and an irreducible form generalized
to higher orders.

Throughout this paper, $N$ is referred to as the total number of particles,
$K$ the total number of cells or grid points, and $p$ the order of multipole
expansions. We may assume $K \sim N/p^2$ for the optimum value,
however, we leave $N$ and $K$ as separate parameters for convenience.

This paper is organized as follows.
In section 2, 
we present the algorithm in detail.
In section 3, we discuss the computational complexity of PMMM
for a given set of parameters, and give a guide to chose the
parameters.
A test of numerical accuracy is carried out in section 4.
Finally in section 5, we discuss a possible parallelization
and hierarchical version of PMMM, and 
applications to classical simulation of 
biomolecular systems and cosmological $N$-body simulations.
The appendix supplies some materials useful for implementation.

\section{Construction}

\subsection{Algorithm in detail}
Consider $N$ particles distributed in $K$ uniform cells.
For each cell, the set of particles inside is known. 
This condition is
achieved in an $\mathcal O(N)$ procedure.
The multipole moments of each cell in $(p+1)^2$ real numbers
are evaluated with (\ref{eq:P2M}).
Let us now express the multipole moments of cell $\bm i = (i_x, i_y, i_z)$
as a vector
$\bm M_{\bm i} \in \mathbb R^{(p+1)^2}$ where $\bm i = (i_x, i_y, i_z) \in \mathbb Z^3$ 
are three dimensional indices of the cell.
The local expansions of each cell 
$\bm L_{\bm i} \in \mathbb R^{(p+1)^2}$
is available in
\begin{equation}
\bm L_{\bm i} = \sum_{\bm j} {\mathsf G}_{\bm i - \bm j} \bm M_{\bm j}.
\label{eq:pmmm}
\end{equation}
Here, ${\mathsf G}_{\bm i - \bm j}$ is a square matrix of size $(p+1)^2$,
which depends on the displacement vector $\bm i - \bm j$
of cells $\bm i$ and $\bm j$,
and also referred to as a Green's function. 
In the summation,  $\bm j$ iterates over all $K$ cells,
and the indices $\bm i - \bm j$ are cyclic.
An explicit form of ${\mathsf G}_{\bm i - \bm j}$ is given by
(\ref{eq:M2L}), (\ref{eq:realmatrix}), and (\ref{eq:greenPBC}),
and it effectively consists of $(2p+1)^2$ real numbers, not $(p+1)^4$ numbers.

The calculation cost of the summation can be reduced from $\mathcal O(K^2)$ to
$\mathcal O(K \log K)$ using a convolution theorem with FFT.
The following gives the final procedure for periodic systems:
\begin{equation}
\begin{aligned}
\{ \widetilde{{\mathsf G}}_{\bm k} \}  &:= {\mathcal F}_{\bm i\to \bm k} \{ {{\mathsf G}}_{\bm i} \},  \\
\{ \widetilde{\bm M}_{\bm k} \}  &:= {\mathcal F}_{\bm i\to \bm k} \{ {\bm M}_{\bm i} \},  \\
\widetilde{\bm L}_{\bm k}        &:= \widetilde{{\mathsf G}}_{\bm k} \widetilde{\bm M}_{\bm k},  \\
\{ {\bm L}_{\bm i} \}            &:= {\mathcal F}_{\bm k\to \bm i}^{-1} \{ \widetilde{\bm L}_{\bm k} \}. 
\end{aligned}
\end{equation}
Here, $\{\,\}$ denotes the set of all $K$ points,
$\mathcal F$ and $\mathcal F^{-1}$ are forward and backward 
discrete Fourier transforms,
tilde is the value in wave space,
and there are wave-number indices $\bm k \in \mathbb Z^3$.
The first line requires $(2p+1)^2$ Fourier transforms of size $K$ which can be 
performed and saved at the beginning of the simulation.
For the second and the fourth equation, we perform $(p+1)^2$ independent
FFTs for each.
And in the third, we perform element-by-element M2L transformations.
Strictly speaking, 
the effective number of points in complex numbers after
the FFT of $K$ real numbers is $K/2$.
And in the M2L transformations
in the wave space, all the real operations
in (\ref{eq:realmatrix}) are turned into complex operations which are
expected to be four times more expensive.
In total, the cost is equivalent to about $2K$ transformations in real numbers.
For three-dimensional open boundary systems, this convolution procedure
need to be performed on $8K$ points.

After the local expansions of each cell are obtained,
the potential of a particle is available in (\ref{eq:L2P}), 
and its gradient in  (\ref{eq:L2L}) and (\ref{eq:phigrad}).

A cutoff for the nearest 27 or 125 cells is expressed in a
mask in Green's function and contributions from the masked
cells are evaluated in direct particle--particle interactions.
An example layout of Green's function is shown in Fig.~\ref{fig:green}
for a two-dimensional open boundary system.
In a periodic system, the value of Green's function for
the closest interactions is not exactly zero,
and has the contributions from mirror images.

\subsection{Extension of the PM method}
At the limit of the spatial order $p=0$, PMMM agrees with the
PM method in the nearest grid point (NGP) mode,
where equation (\ref{eq:pmmm}) reduces to a scalar equation
\begin{equation}
\Phi_{\bm i} = \sum_{\bm j} G_{\bm i - \bm j} \rho_{\bm j},
\end{equation}
with a discreet scalar potential and density field
$\Phi_{\bm i}$ and $\rho_{\bm j}$, and Green's function $G_{\bm i - \bm j}$.
From this baseline, PM and PMMM increase the spatial order in different ways.
The PM method increases the order with a diffusive interaction
between  a particle and its nearest $(p+1)^3$ grid points.
For $p=0,1,2$, they are called NGP, CIC (cloud in cell), and 
TSC (triangular shaped cloud) mode \citep{Hockney:1988:CSU},
and higher order generalization is given in B-spline functions.
In PMMM, a particle interacts with the nearest grid point
while a grid point holds multiple information in $(p+1)^2$ terms.
From the PM method, scalar density is extended to multipole moments,
scalar potential is extended to local expansions,
and the scalar Green's function is extended to matrix form.

The particle--particle interaction takes different forms.
In the PM series, it is sometimes omitted to give a mesh softening
(pure PM), or it has a cutoff function so as to make the total force
Newtonian (PPPM and PME).
In PMMM, the short range cutoff of the mesh part is expressed 
by the nearest cells as a mask of the Green's function, and the 
particle-particle interaction takes a pure Newtonian form.

\subsection{Optimization from FMM}
At the level of the smallest cells, PMMM agrees with FMM.
The relative positions and charges (masses) of particles are assigned to
the cell center as multipole moments, and the potentials are assigned back
to the particles from local expansions of the cell.
Differences exist in the process to compute the local expansions
from the multipole moments of all the other cells.
For the number of cells $K$, a simple summation takes
$K(K-1)$ multipole-to-local (M2L) transformations.
FMM exploits a hierarchical octree structure for reduction to
$\mathcal O(K)$.
Instead, PMMM employs a uniform mesh structure and the transformations
are accelerated by the fast convolution theorem using FFT. 
It only requires $K$ transformations, associated with an
extra cost of FFT which is $\mathcal O(K\log K)$.
However, a factor 8 overhead exists for an isolated system.

The $\mathcal O(K\log K)$ scaling does not immediately mean it is slower than
that of $\mathcal O(K)$ in practical cases. The $\mathcal O(K)$ method tends to have a 
relatively large factor, about 189 to 875, depending on the cell separation criterion.
In the $\mathcal O(K\log K)$ scheme, the number of transformations
remains constant, $K$ or $8K$ irrespective of the criterion. 
In both schemes, it is common that the contributions from the nearest
cells are evaluated through direct particle-particle interactions.

\section{Computational and memory complexity}
\subsection{Order estimation}

\begin{enumerate}
\item Each particle interacts with $(p+1)^2$ coefficients of the
nearest grid point. This part is $\mathcal O(p^2 N)$.
\item The forward and backward FFTs of size $K$ are performed for 
$(p+1)^2$ independent terms,
and is $\mathcal O(p^2 K \log K)$.
\item The M2L transformations 
$\mathbb C^{(p+1)^2} \to \mathbb C^{(p+1)^2}$
are performed on the number of $K$ reciprocal grids,
and is $\mathcal O(p^4 K)$.
\item Each particle interacts with $\mathcal O(N/K)$ particles in nearby cells,
with a multiplying factor $27$ or $125$. In total, this part is $\mathcal O(N^2/K)$.
\item The memory demand is $\mathcal O(N + p^2 K)$ including the transformation
matrices.
\end{enumerate}
If we set the parameter $K \propto N/p^2$, the total computational cost becomes
either $\mathcal O(p^2 N)$ or $\mathcal O(N \log (N/p^2))$, and the memory demand
$\mathcal O(N)$.
In this article, we follow the original $\mathcal O(p^4)$ transformation method
by \citep{Greengard:1988:REPb}, however, a possible reduction to $\mathcal O(p^2 \log p)$
of this part is discussed in \S \ref{sec:logp}.

\subsection{Choice of parameters}
\begin{figure}[htpd]
\centering
\begin{subfigure}[h]{.5\linewidth}
  \includegraphics[clip,width=1.0\linewidth]{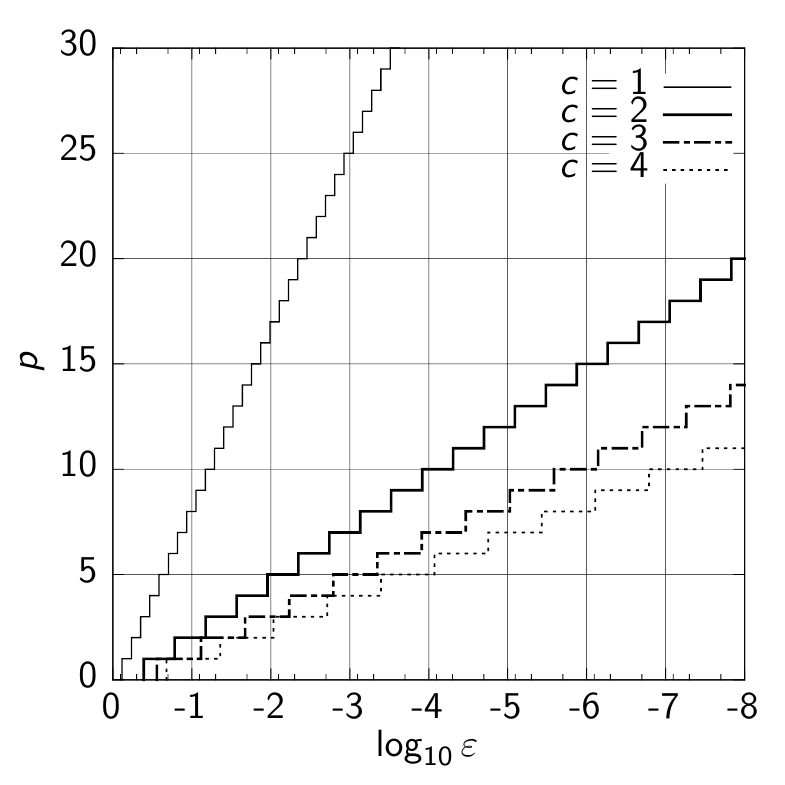}%
\end{subfigure}%
\begin{subfigure}[h]{.5\linewidth}
  \includegraphics[clip,width=1.0\linewidth]{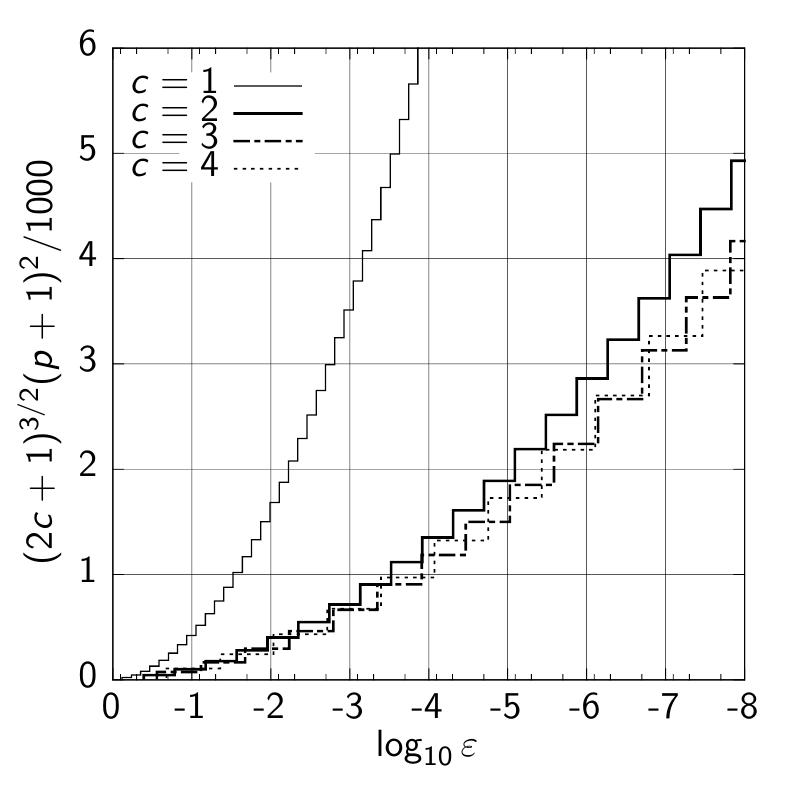}%
\end{subfigure}%
\caption{
	The order of expansion $p$ (left) and the cost scaling
	$(2c+1)^{3/2} (p+1)^2$ (right) for given error tolerances $\varepsilon$,
	for $c=1,2,3,4$.
}
\label{fig:cost}
\end{figure}

We put the minimum cell separation as $c \ (\ge 1)$,
and try to find the optimum value of the parameters including
$K$ and $p$, for given error tolerance.
Let us write the cost of the short range particle--particle interactions 
and the long range particle-mesh interactions as
\[
C_{\rm PP} (2c+1)^3 N^2/K, \quad\text{and}\quad C_{\rm PM} K (p+1)^4.
\]
Here, we assume that the $\mathcal O(p^4 K)$ translation part costs
more than the $\mathcal O(p^2 K \log K)$ FFT part.
The factor 8 overhead for the open boundary case can be included
to the coefficient $C_{\rm PM}$.
The balancing point of these two is given by
\[
K = \sqrt{C_{\rm PP} / C_{\rm PM}} (2c+1)^{3/2} N / (p+1)^2, \nonumber
\]
with the resulting total cost
\[
2 \sqrt{C_{\rm PP} C_{\rm PM}} N \cdot (2c+1)^{3/2} (p+1)^2. \nonumber
\]
The error in the worst case is estimated by
\begin{equation}
\label{eq:errest}
\varepsilon = \left( \frac{\sqrt3 / 2}{(c+1) - \sqrt3 / 2} \right)^{p+1}.
\end{equation}
This gives scalings $0.76^p$, $0.41^p$ for $c=1,2$ 
\citep{Greengard:1988:REPa} and $0.28^p$, $0.21^p$  for $c=3,4$.
The order of expansion $p$ and the cost scaling $(2c+1)^{3/2} (p+1)^2$
for the given tolerance $\varepsilon$ are plotted in Fig.~\ref{fig:cost}.
A large offset in the efficiency exists between $c=1$ and $c=2$,
and from $c=2$, they all behave similarly.
Thus, $c=2$ (125 cells cutoff) seems satisfactory in most cases, though
larger cutoff can be considered  when  $p \ge 10$ is needed for $c=2$
(i.e. $\varepsilon < 10^{-4}$).

\section{Numerical test}
Even the error behavior of PMMM can be expected
--- 
it should be the same level or slightly better than that of FMM
because the result is  mathematically equivalent to the 
summation of $K^2$ transformations of all cells
---
we carried out minimum numerical tests for the verification
of the scheme.
The potential and its gradient obtained from an open boundary PMMM
were compared with those from the $\mathcal O(N^2)$ direct summation,
and those of a periodic bondary PMMM were compared with the Ewald summation.
The test code was implemented in C++ with double precision
arithmetics, and we used the FFTW 3.3.3 
\footnote{
	\url{http://www.fftw.org}
}
library for the three-dimensional real-to-complex and complex-to-real
Fourier transforms.

For the test condition, we randomly distributed $N=16,384$ particles
in a unit box, and the box was split into $K=8^3$ cells, while the
FFT and M2L translations in the wave space were performed on
$16^3$ cells for the open boundary condition.
The charges of the particles were also randomly distributed in
the range $[0, 1/N)$, but were later subtracted by the average value
to make the total charge of the system zero.
We tested both the 27 cells cutoff (one cell for the minimum separation) 
and the 125 cells cutoff (two cells separation) cases.

\begin{figure}[htpd]
\centering
\begin{subfigure}[h]{.5\linewidth}
  \includegraphics[clip,width=1.0\linewidth]{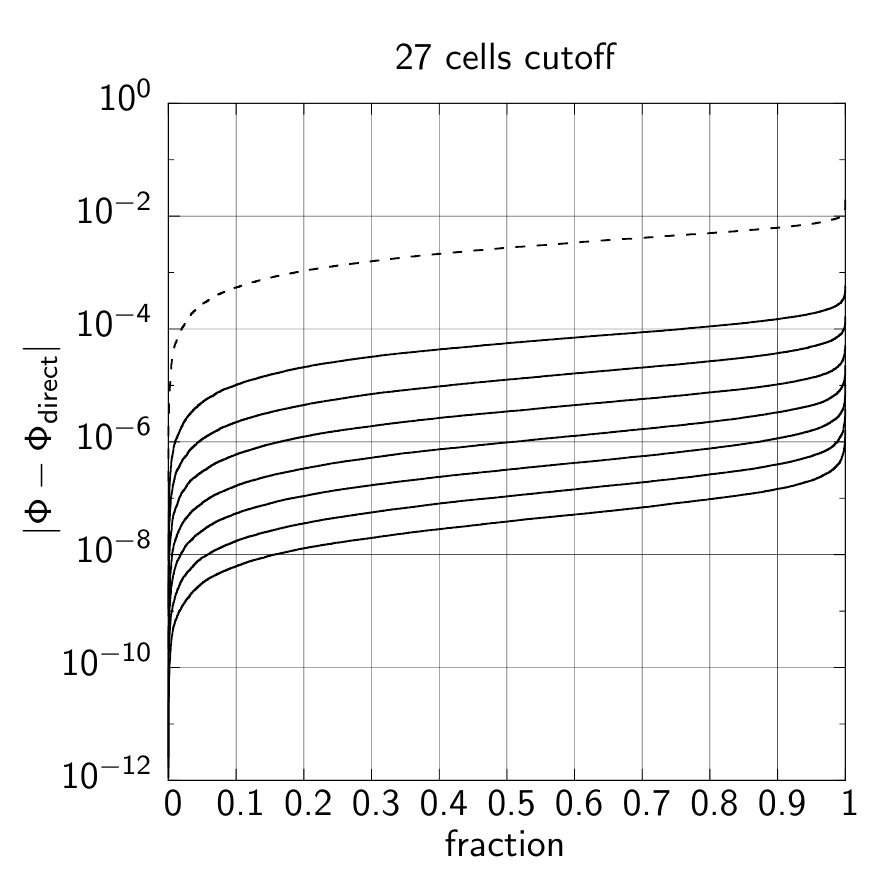}%
\end{subfigure}%
\begin{subfigure}[h]{.5\linewidth}
  \includegraphics[clip,width=1.0\linewidth]{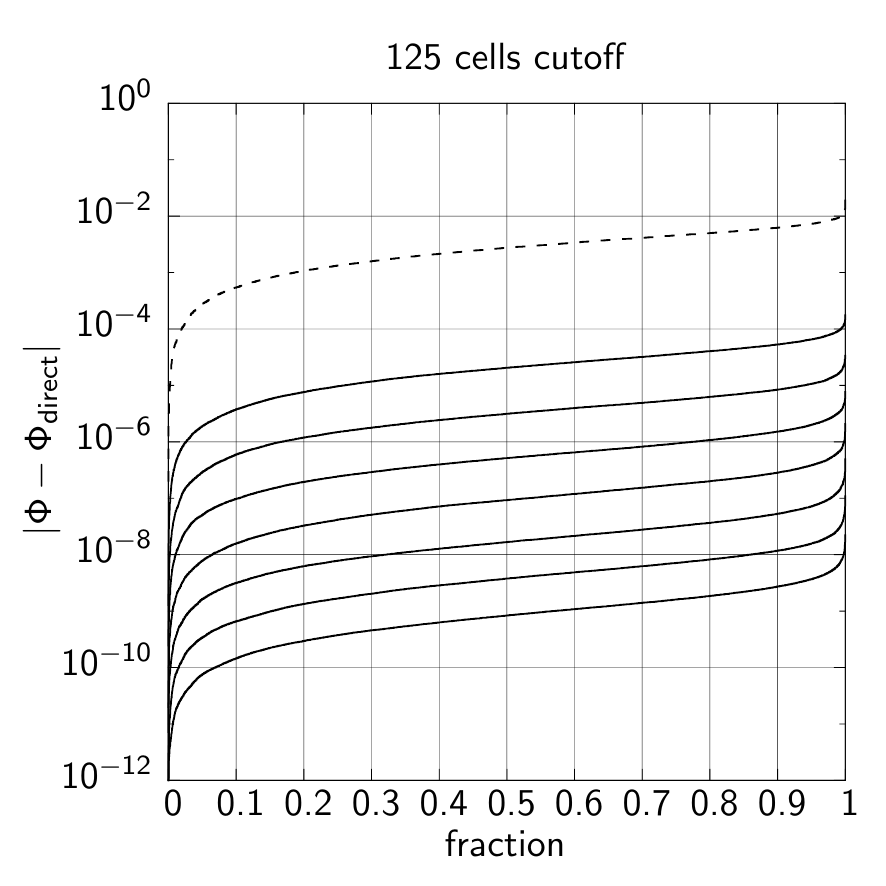}%
\end{subfigure}%
\\
\begin{subfigure}[h]{.5\linewidth}
  \includegraphics[clip,width=1.0\linewidth]{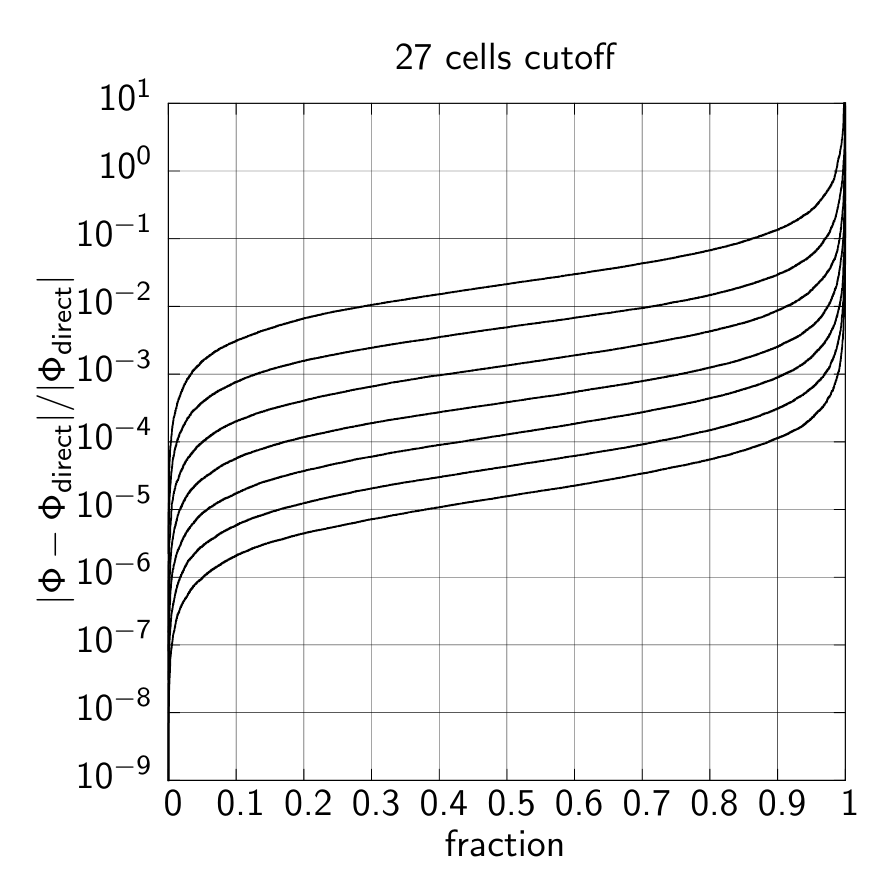}%
\end{subfigure}%
\begin{subfigure}[h]{.5\linewidth}
  \includegraphics[clip,width=1.0\linewidth]{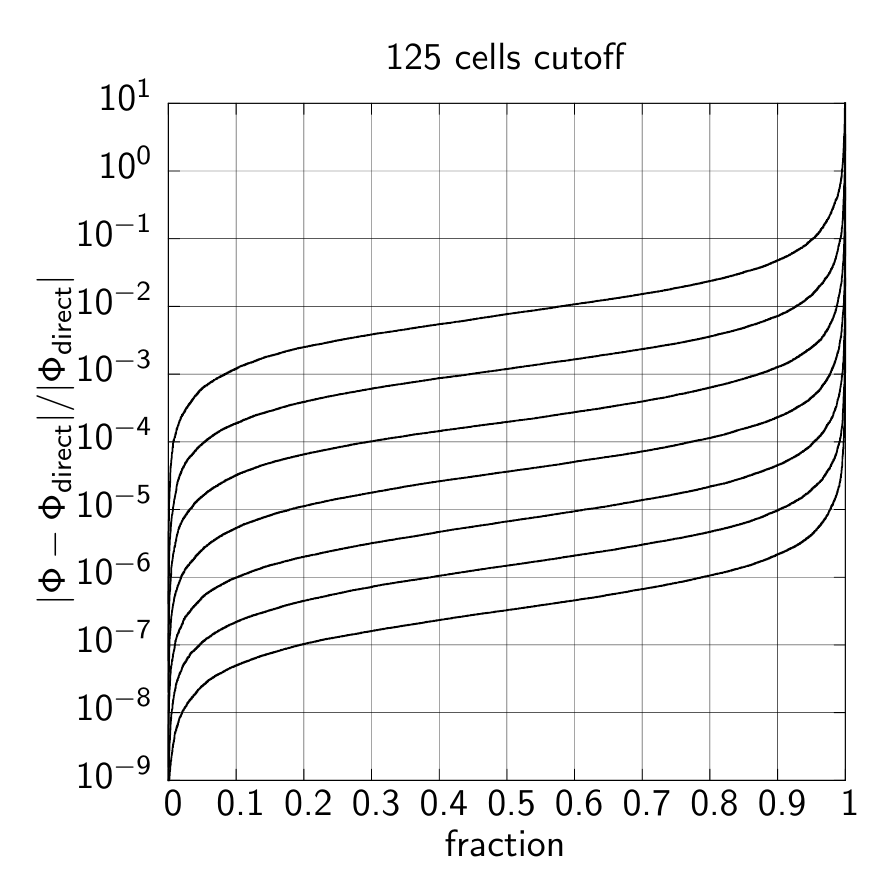}%
\end{subfigure}%
\caption{
	Cumulative plots of the error in potential,
	absolute (top) and relative (bottom) error 
	for 27 cells cutoff (left) and 125 cells cutoff (right),
	for the order of expansion $1 \le p \le 7$ (upper curve
	to lower curve).
	The cumulative distribution of potential itself is plotted in 
	the dashed curve.
}
\label{fig:errpot}
\end{figure}

\begin{figure}[htpd]
\centering
\begin{subfigure}[h]{.5\linewidth}
  \includegraphics[clip,width=1.0\linewidth]{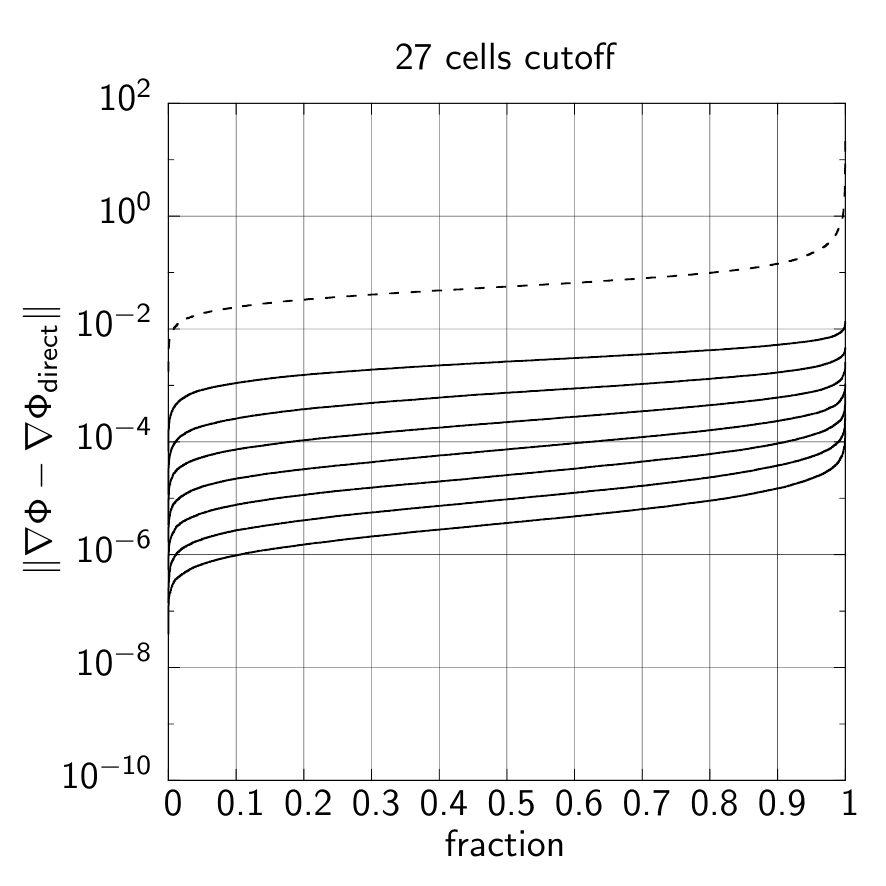}%
\end{subfigure}%
\begin{subfigure}[h]{.5\linewidth}
  \includegraphics[clip,width=1.0\linewidth]{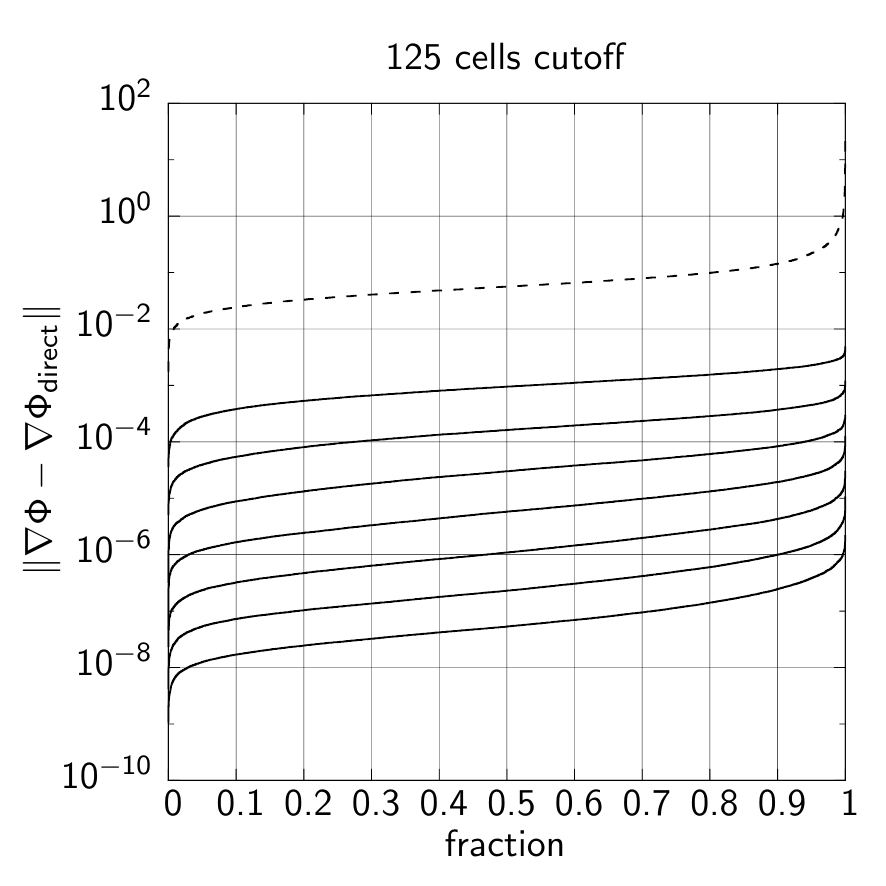}%
\end{subfigure}%
\\
\begin{subfigure}[h]{.5\linewidth}
  \includegraphics[clip,width=1.0\linewidth]{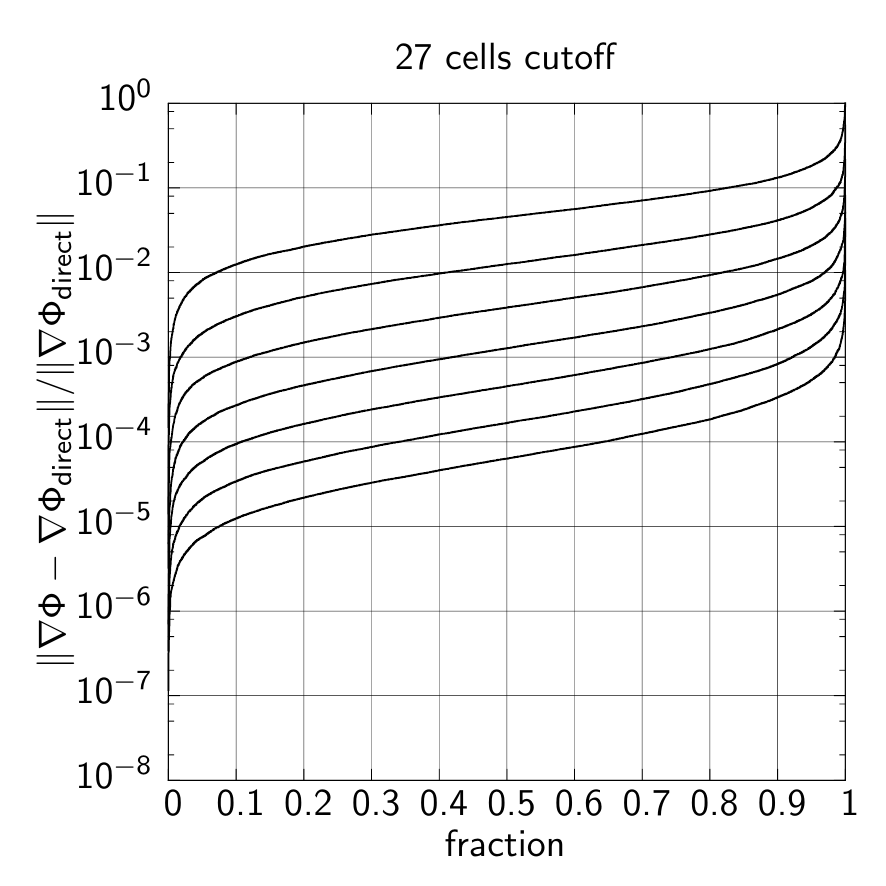}%
\end{subfigure}%
\begin{subfigure}[h]{.5\linewidth}
  \includegraphics[clip,width=1.0\linewidth]{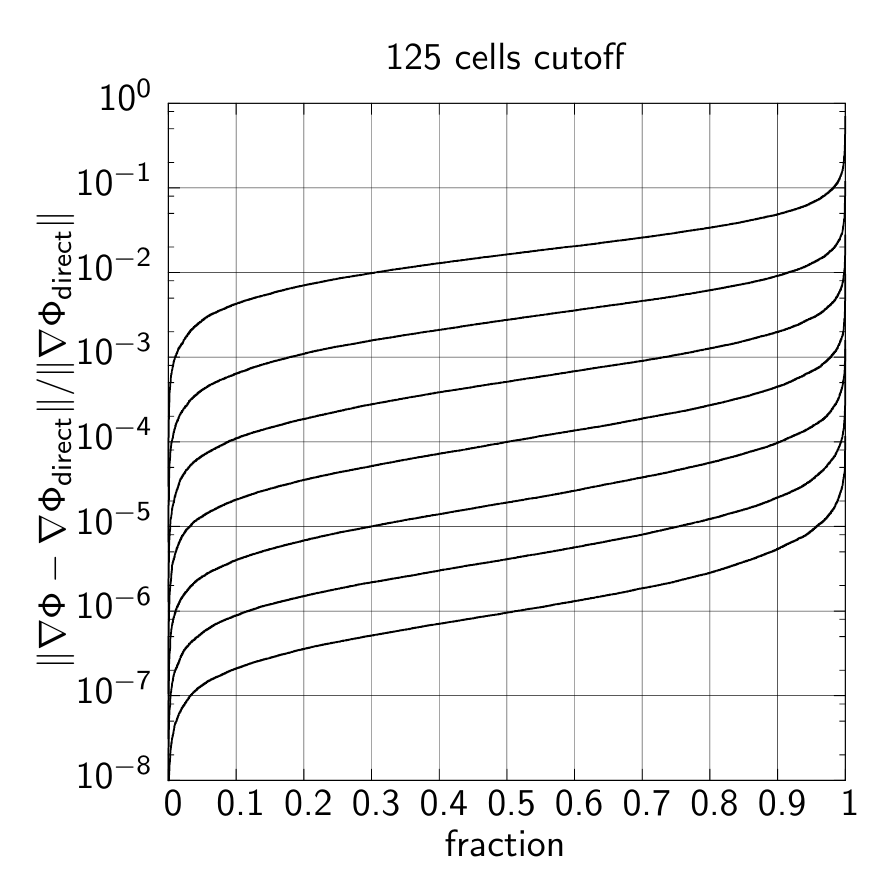}%
\end{subfigure}%
\caption{
	Same as Fig.~\ref{fig:errpot}, but for the gradient of potential $\nabla \Phi$.
}
\label{fig:errgrad}
\end{figure}

Fig.~\ref{fig:errpot} shows cumulative distributions of absolute and 
relative error in potential for the expansion order $1 \le p \le 7$,
in open boundary calculations.
To make it a dimension free comparison,
the distribution of the absolute value
of the potential itself is plotted as a reference.
Fig.~\ref{fig:errgrad} is essentially the same, but the norm of the error
in the gradient of the potential is plotted.
All of the plots represent well what we can  expect from the multipole theory.
The error decreases by a ratio as $p$ increases, and the larger cutoff improves
the convergence, though the scaling looks even better than the
estimation in (\ref{eq:errest}).

\begin{figure}[htpd]
\centering
\begin{subfigure}[h]{.5\linewidth}
  \includegraphics[clip,width=1.0\linewidth]{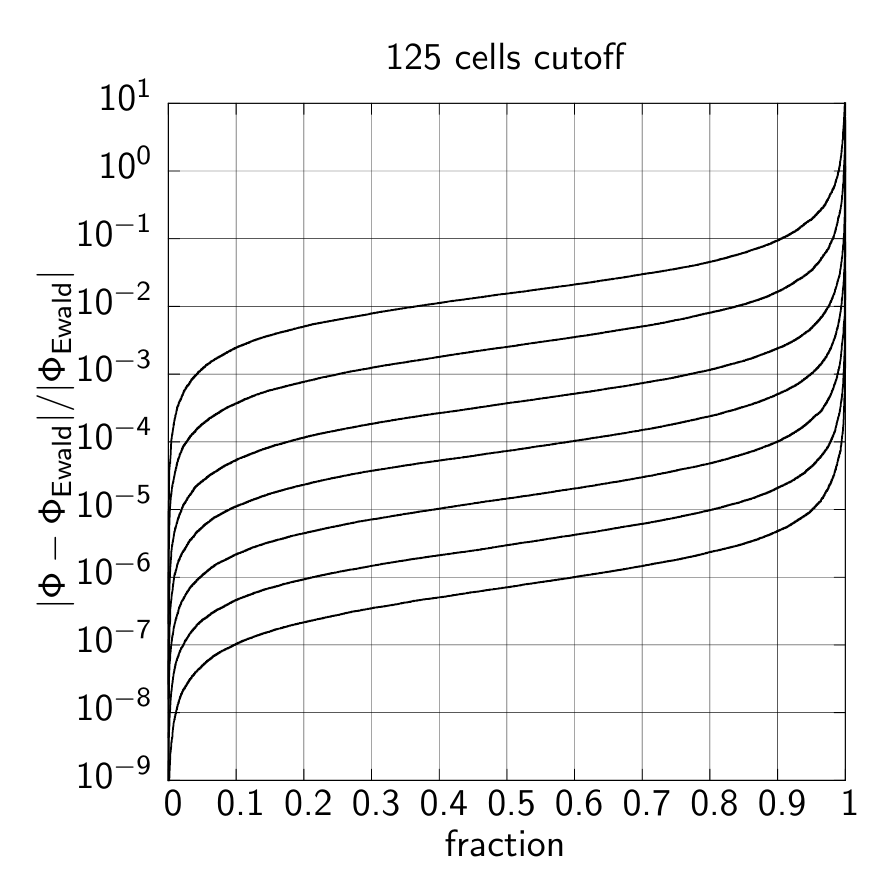}%
\end{subfigure}%
\begin{subfigure}[h]{.5\linewidth}
  \includegraphics[clip,width=1.0\linewidth]{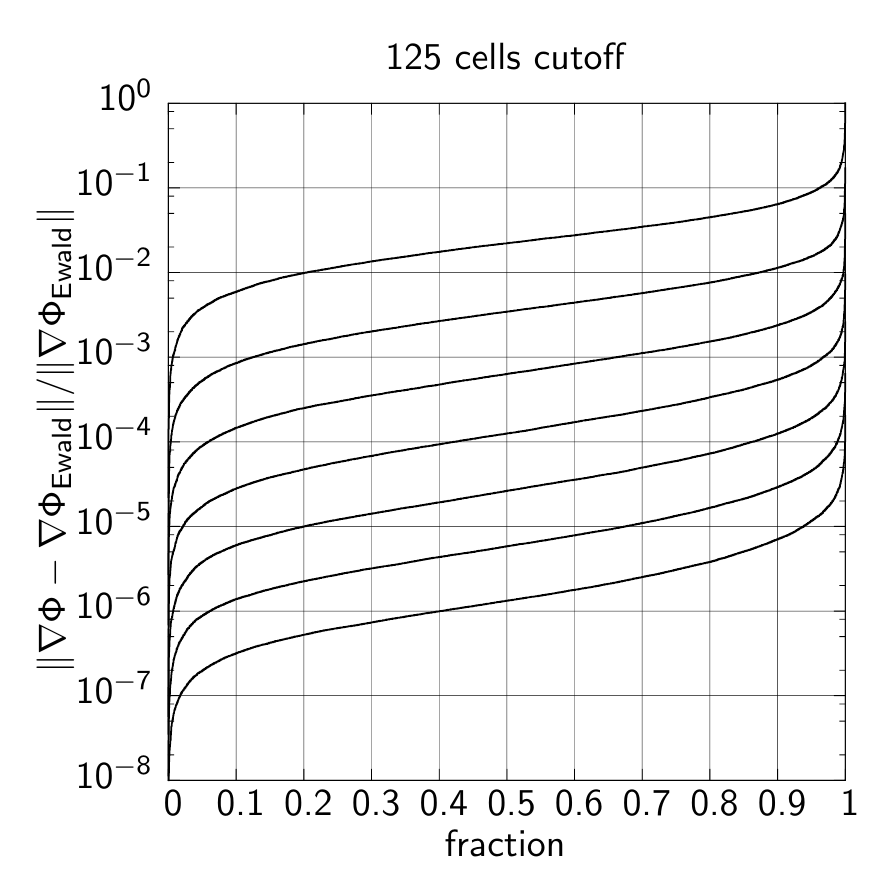}%
\end{subfigure}%
\caption{
	Relative error in potential (left) and its gradient (right)
	in a periodic boundary system, for $1 \le p \le 7$.
}
\label{fig:pbc}
\end{figure}

Errors in a periodic boundary system with the same particle distribution
above are plotted in Fig. \ref{fig:pbc}.
The potential and its gradient were evaluated  as described
in \ref{sec:greenpbc} and compared with those from the Ewald summation.
The error behavior is common to that of an open boundary case.

\section{Discussion}
\subsection{Parallelization}
Parallelization of PMMM for distributed memory machines is a 
straightforward task and the communication pattern is totally regular.
As an extreme case, we consider using the number of processors 
equal to the number of cells $K$ and assume that each processor is
responsible for one cell and the particles contained in it.
The interactions between particles and cell, 
particle-to-multipole (P2M) and local-to-particle (L2P)
are totally local and parallel.
The cell--cell interactions, M2L translations in the wave space 
after the FFT are
also local and parallel, with a distributed Green's function 
(transformation matrices).
Half the number of processors can join this part if we employ
real-to-complex and complex-to-real transforms.
The particle--particle interactions of nearby cells need to be
solved by communications with neighbor processors, and this part is simple.
Thus, almost all of the efforts will be
focused on the efficient communication for the forward and backward FFT part
for the $(p+1)^2$ independent terms.

One choice is to use $(p+1)^2$ processors for the FFT part where
each FFT processor gathers/scatters the elements from/to
all other processors.
This can be easily written with nonblocking collective operations
supported as a new feature in the message passing interface (MPI) version 3.0 \citep{mpi-3.0}.
The communication size that one cell exports/imports during a
step is minimum and usually smaller than that of an octree based FMM;
sending and receiving the $(p+1)^2$ terms for the forward and backward FFTs.
However, in this naive algorithm,
heavy communication traffic can concentrate on the processor
that performs FFT.

In more practical cases, we may use the number of processors
less than the number of cells $K$, and use more than one
processors for each FFT. Still the discussion above is unchanged
that all we have to take care of is the parallel FFT part for the
distributed coefficients.

An implementation for a graphics processing unit (GPU) 
would be easy by assigning each cell for each thread,
and $(p+1)^2$ thread-blocks for the multiple FFT tasks.

We remark that a parallel version of FFTM already exists \citep{journals/tcad/OngLL04}.

\subsection{Hierarchical treatment}
\label{sec:hierarchical}
It might be a common argument from the computational science side,
that an FFT and uniform grid based scheme wastes the {\it local} nature
of the system.
In principle, distant interactions can be performed in large scale cells.
Moreover, the contribution from the higher order moment is more local ---
equation (\ref{eq:M2L}) expresses that the contribution from the
$\lambda$-th order multipole moment to the $\ell$-th order local
expansion decays as $1/r^{\ell+\lambda+1}$.

Here, we discuss the case that we use a global coarse mesh
and local fine mesh instead of a uniform mesh, with
$K = K_{\rm coarse} K_{\rm fine}$.
First, PMMM is directly applied on the coarse mesh resulting in
the cost of $K_{\rm coarse}$ transformations in total,
and the contribution from/to the nearest 27 or 125 coarse
cells (including the self cell) are left over.
Then a coarse cell is refined to $K_{\rm fine}$ 
(e.g. $4^3$) fine cells. From a set of the multipole moments
of $K_{\rm fine}$ cells, local expansions of
$27 K_{\rm fine}$ or $125 K_{\rm fine}$ cells can be evaluated
in a convolution form.
Fig.~\ref{fig:refine} illustrates a possible layout for
 the two-dimensional case.
Including the buffer zone for the aliasing effect, one fine cell
costs 64 or 216 transformations
\footnote{
	$64 = (3+1)^3$, $216 = (5+1)^3$, $189 = 6^3-3^3$, $875 = 10^3 - 5^3$.
}
irrespective of the refinement size $K_{\rm fine}$.
These numbers are slightly better than those of  the original FMM,
189 or 875,
where the extra cost comes from the asymmetry of the octree structure.
However, they are much more expensive than the uniform mesh case.
This hierarchical treatment relaxes the demand for the
global network bandwidth in exchange for increased local
computations.

For the refinement factor $K_{\rm fine}$, any cubic number can
be used, like $4^3$, $8^3$ or even $6^3$, etc.
Thus, the hierarchy can be quite shallow compared with the octree
structure, and two levels are usually enough.
As a bonus of the hierarchical treatment, an $\mathcal O(N)$ scaling is recovered
from the $\mathcal O(N \log N)$, where the cost associated with
the fine cells dominates.

\subsection{Application to molecular dynamics (MD)}
The most significant bottleneck
in the simulations of bimolecular systems is,
especially for the distributed memory parallel computers of today,   
in solving the global electrostatic force from charged atoms.
Here, we discuss the use of PMMM for such simulations
with typical parameters.
We consider a simulation box with a size $(100\text{\AA})^3$,
and number density $\sim 0.1$ atom/\AA$^3$, i.e. $N=10^5$.
We take the cell size as {5\AA}  and 125 cutoff neighbors,
to make the  cutoff length 10\AA. This is a typical value of the
cutoff length for the Lennard-Jones potential.
In the PME method
\citep{Darden:1993:PME} 
and its variants, smooth particle mesh Ewald \citep[SPME,][]{essmann95a}
and Gaussian split Ewald  \citep[GSE,][]{Shan2005:GSE}
being the most popular solvers in this field, 
a typical grid spacing in the above case is 1\AA.
Hence, the comparison is between single FFT of size $100^3$ 
or multiple $(p+1)^2$ FFTs of size $20^3$.
At $p=7$, the total number of coefficients on the grid points
becomes about half of PME, keeping the accuracy at a reasonable level. 
This helps for reducing the FFT overhead.
Moreover, lower precision arithmetics can be used for the
higher order coefficients;
e.g. starting from a 24-bit fixed point number at $\ell=0$,
cutting every 3-bit for an increment of $\ell$ results into
only 612 bits per cell with $p=7$
\footnote{
	An $\ell$-th order term has $3 \cdot (p+1-\ell)$ bits and
	there are $(2\ell+1)$ terms.
	The total number of bits is
	${3 \cdot \sum_{\ell=0}^p (2\ell+1)(p+1-\ell)} = {3 \cdot (p+1)(p+2)(2p+3)/6}$.
}.
A bit-level compression technique for higher order derivatives
is frequently used in special-purpose hardware
\citep{Makino:2003:GMP}.

In addition, the number of coefficients that a particle interacts with 
is not a negligible factor for the efficiency.
In the SPME method, a particle usually interacts with 216 grid points for $p=5$,
while 64 terms of the nearest grid point are needed in PMMM with $p=7$.
Furthermore, PMMM has little difficulty in the charge assignment part
in a multi-thread environment because a particle only interacts with its
nearest grid point.

In future work, we will address the question of the sufficient
value of $p$ in practical simulations, and compare the accuracy
and efficiency of PMMM with PME series.
Finally, we emphasize the benefit of the relatively simple form  of the 
particle--mesh interaction without diffusion and particle--particle
interaction without a cutoff function.

\subsection{Application to cosmological N-body simulations}
PMMM has the same dilemma as PPPM when applied directly to
cosmological $N$-body simulations.
At the late stage of the simulation after structure formation,
a high density contrast starts to require either increased
particle--particle interactions or finer mesh size.
Thus, a relatively young TreePM method \citep{Bagla:2002:TCC} 
which replaces the PP part of PPPM with an adaptive octree algorithm 
\citep{Barnes:1986:HFC} has become the most popular solver
in this field, due to its simplicity and efficiency.
Exactly the same approach can be applied for PMMM, where the
global force field is solved by PMMM with the periodic boundary
condition. For the contribution from the 27 or 125 cutoff cells,
either multipole-to-particle (M2P) based Barnes \& Hut type tree method
or M2L based FMM on adaptive octree structure can be applied. 

The practical benefit of such combinations is the simplicity
of both the code structure and accuracy control.
The adaptive octree structure can be started from each cell
of PMMM when it is needed, and constructed locally.
The accuracy of the mesh part is simply controlled with 
one parameter $p$.
Since the PP part has no cutoff function,
it is easier to introduce higher order treecode than the
center-of-mass approximation ($p=1$), which most of TreePM
implementations employ \citep{Springel:2005:CSC, Ishiyama:2009:GMP}.

\subsection{Another use of FFT convolution}
\label{sec:logp}
Historically, the use of fast convolution with FFT was suggested
at an early time by \cite{Greengard:1988:EIF} to reduce the complexity
of each M2L translation from $\mathcal O(p^4)$ to $\mathcal O(p^2 \log p)$.
Later, a full study was carried out by \cite{Elliott:1996:FFT} including
the treatment of numerical instabilities.
A two-dimensional convolution form of an M2L transformation is visualized
in Fig.~\ref{fig:convmtol}.
This approach should not be confused with the other where the
FFT convolution is performed on uniform grids 
as in \citet{Shimada:1993:ECC}, \citet{Ong:2003:FAT}, and this work,
although the approaches may be categorized by the keywords
`FFT based FMM'.
Ultimately, these two methods can be combined through a five-dimensional
FFT resulting in the total cost $\mathcal O(p^2 K \log (p^2 K))$.
Still it is not clear whether an application exists
that requires an extremely high order of expansion.

\section*{Acknowledgement}
The author thanks Dr.~Rio Yokota of King Abdullah University of Science and Technology
for his useful comments on FMM.
The author also thanks Dr.~Yousuke Ohno of RIKEN QBiC for his comments
on biomolecular simulations.

A minimum implementation of PMMM can be obtained at \\
\url{https://github.com/nitadori/PMMM}.

\appendix
\setcounter{figure}{0}

\section{Implementation note for FMM}
The original formulation of the three-dimensional FMM for 
$1/r$ potential based on spherical harmonics \citep{Greengard:1988:REPb} 
seems to be a bit complicated and
not straightforward for implementers.
In this Appendix, we provide all the equations needed for the implementation.
Strict proofs are mostly omitted, see 
\cite{Epton:1995:MTT:210669.210688}
and
\cite{Gelderen98theshift}
for full descriptions.

\subsection{Solid harmonics}
A simple formulation of the multipole transformations is
established with two solid harmonics functions \citep{Wang:1996:EFM}.
We first prepare definitions of `regular' and `singular' 
solid harmonics base functions in spherical coordinates,
\begin{align}
R_\ell^m (r,\theta,\phi) &=   
    \frac{r^\ell}{(\ell+m)!} P_\ell^{m}(\cos\theta) e^{im\phi}, \\
S_\ell^m (r,\theta,\phi) &=
    (-1)^{\ell+m} \frac{(\ell-m)!}{r^{\ell+1}} P_\ell^{m}(\cos\theta) e^{im\phi},
\end{align}
for $0 \le m \le \ell$. 
Extensions for negative $m$ is given by the conjugate relation
$R_\ell^{-m} = (-1)^m [{R_\ell^{m}}]^*$
and
$S_\ell^{-m} = (-1)^m [{S_\ell^{m}}]^*$.
A sign factor $(-1)^{\ell+m}$ is included in the definition of
$S_\ell^m$ for simplicity of later formulation.
Here, $P_\ell^m$ is the associated Legendre polynomial which follows the definition,
\begin{equation}
\begin{split}
P_\ell^m(\cos\theta) 
	&= (-\sin\theta)^m \frac{d^m}{(d\cos\theta)^m} P_\ell (\cos\theta) \\
	&= \frac{(-\sin\theta)^m}{2^\ell \ell!}  \frac{d^{\ell+m}}{(d\cos\theta)^{\ell+m}} 
	\left[ (\cos\theta)^2 - 1 \right]^\ell.
\end{split}
\end{equation}
Note that another definition exists for the factor $(-1)^m$.

These functions represent a general solution of the Laplace equation 
in spherical coordinates
$\nabla^2 \Phi(r, \theta, \phi) = 0$, where
\begin{equation}
\Phi(r, \theta, \phi) = \sum_{\ell=0}^{\infty} \sum_{m=-\ell}^{\ell}
\bigl[
	M_\ell^m S_\ell^{-m} (r, \theta, \phi) +
	L_\ell^m R_\ell^m (r, \theta, \phi)
\bigr],
\end{equation}
for the outer (former term) and inner (latter term) solutions.
The coefficients $M_\ell^m$ and $L_\ell^m$ are referred to as
{\itshape multipole moments} and {\itshape local expansions}, respectively.

Some low order Cartesian expressions of these base functions 
are listed in table \ref{tab:RSlm}.
It turns out that both functions have simple Cartesian forms
and hereafter we employ the notation
$R_\ell^m(\bm r) = R_\ell^m(r, \theta, \phi) = R_\ell^m(x, y, z)$.

\begin{table}[htbp]
\caption{$R_\ell^m$ and $S_\ell^m$ in low orders.}
\label{tab:RSlm}
\centering
\begin{tabular}{ccc}
\hline
$(l,m)$ & $R_\ell^m$ & $S_\ell^m$ \\
\hline
$(0, \phantom\pm 0)$ & $1$ &  $1/r$ \\
$(1, \phantom\pm 0)$ & $z$ &  $-{z}/{r^3}$ \\
$(1,\pm 1)$ & $-(\pm x+iy)/2$ &  $-(\pm x+iy)/{r^3}$ \\
$(2, \phantom\pm 0)$ & $ (3z^2-r^2)/4$ &  $(3z^2-r^2)/r^5$ \\
$(2,\pm 1)$ & $ -z(\pm x+iy)/2$ &  $3z(\pm x+iy)/r^5$ \\
$(2,\pm 2)$ & $ (\pm x+iy)^2/8$ &  $3(\pm x+iy)^2/r^5$ \\
\hline
\end{tabular}
\end{table}

\subsection{Ladder operators}
This section is for the preparation of the derivations
of transformation relations, and could be skipped when
the interest is only in the results.

We define the following three {\itshape ladder operators}
\begin{equation}
\partial_+  =  \dfrac{\partial}{\partial x} + i \dfrac{\partial}{\partial y}, \quad
\partial_z  =  \dfrac{\partial}{\partial z}, \quad
\partial_-  = -\dfrac{\partial}{\partial x} + i \dfrac{\partial}{\partial y},
\end{equation}
where the base functions satisfy
\begin{equation}
\begin{pmatrix}
  \partial_+ \\
  \partial_{\mathrlap{z}\phantom{+}} \\
  \partial_- 
\end{pmatrix}
R_\ell^{m}(\bm r)
=
\begin{pmatrix}
  R_{\ell-1}^{m+1}(\bm r) \\
  R_{\ell-1}^{m\hphantom{+1}}(\bm r) \\
  R_{\ell-1}^{m-1}(\bm r)
\end{pmatrix}
,\quad
\begin{pmatrix}
  \partial_+ \\
  \partial_{\mathrlap{z}\phantom{+}} \\
  \partial_- 
\end{pmatrix}
S_\ell^{m}(\bm r)
=
\begin{pmatrix}
  S_{\ell+1}^{m+1}(\bm r) \\
  S_{\ell+1}^{m\hphantom{+1}}(\bm r) \\
  S_{\ell+1}^{m-1}(\bm r)
\end{pmatrix}
.
\end{equation}
When the indices of $R_\ell^m$ fall outside the range from
$0\le|m|\le\ell$, they merely become zero.
The Laplace equation is confirmed with
$\partial_+ \partial_- [\,] = {\partial_z}^2 [\,]$,
where $[\,]$ is either $R_\ell^m$ or $S_\ell^m$ which is
one of the solutions
\footnote{
	This explains that, if we define 
	$\partial_- = \frac{\partial}{\partial x} - i \frac{\partial}{\partial y}$
	and $Y_\ell^{-m} = [Y_\ell^m]^*$ (without the factor $(-1)^m$),
	the transformation formulae tend to have unwieldy sign factors.
}
.

From these operators, we compose
\begin{equation}
\mathcal D_\ell^{\pm|m|} = 
	(\partial_\pm)^{|m|} (\partial_z)^{\ell-|m|},
\end{equation}
to make 
$\mathcal D_\ell^m R_\lambda^\mu = R_{\lambda-\ell}^{\mu+m}$
and
$\mathcal D_\ell^m S_\lambda^\mu = S_{\lambda+\ell}^{\mu+m}$.
By remembering $R_0^{\, 0} = 1$ and $S_0^{\,0} = 1/r$,
we have
$\mathcal D_\ell^{-m} R_\ell^m = 1$
and
$\mathcal D_\ell^m (1/r) = S_\ell^m$.

Let us examine the relation
\begin{equation}
\label{eq:kronecker}
\left[\mathcal D_\ell^{-m} R_\lambda^\mu (\bm r)\right]_{\bm r 
	= \bm 0} = \delta_{\ell\lambda} \delta_{m\mu},
\end{equation}
with $\delta_{ij}$ the Kronecker delta.
When $\ell \neq \lambda$ or $m \neq \mu$, 
$\mathcal D_\ell^{-m} R_\lambda^\mu (\bm r)$ becomes just $0$ or
a homogeneous polynomial of Cartesian coordinates,
which  vanishes with the substitution $\bm r = \bm 0$.
Then we consider a vacuum potential field around $\bm r_L$
which is expressed in local expansions
\begin{equation}
\Phi(\bm r) = \sum_{\lambda=0}^\infty \sum_{\mu=-\lambda}^{\lambda}
  L_\lambda^\mu R_\lambda^\mu (\bm r - \bm r_L).
\end{equation}
By applying (\ref{eq:kronecker}),
we can extract the coefficient $L_\ell^m$ as
\begin{equation}
\label{eq:extractL}
L_\ell^m = \left[\mathcal D_\ell^{-m} \Phi(\bm r)\right]_{\bm r = \bm r_L}.
\end{equation}
This enables us to obtain a set of local expansion coefficients
at an arbitrary vacuum point of a given potential field
$\Phi(\bm r)$.

Now, a Taylor expansion of a function $\Phi(\bm r)$ that 
satisfies the Laplace equation can be written as
\begin{equation}
\Phi(\bm r + \varDelta \bm r) = 
	\sum_{\lambda=0}^{\infty} \sum_{\mu=-\lambda}^{\lambda}
	\left[\mathcal D_\lambda^{-\mu} \Phi(\bm r)\right]
	R_\lambda^\mu(\varDelta \bm r).
\end{equation}
The expansion is valid as long as the sphere of expansion
is vacuum.
As the base function $R_\ell^m (\bm r)$ or $S_\ell^m (\bm r)$ is a
solution of the Laplace equation, both of them can be expanded
in the same way, yielding addition theorems of solid harmonics,
\begin{align}
\label{eq:additionR}
R_\ell^m (\bm r + \varDelta \bm r) &= 
	\sum_{\lambda=0}^{\infty} \sum_{\mu=-\lambda}^{\lambda}
	R_{\ell-\lambda}^{m-\mu} (\bm r)
	R_\lambda^\mu(\varDelta \bm r) 
	, \\
\label{eq:additionS}
S_\ell^{-m} (\bm r + \varDelta \bm r) &= 
	\sum_{\lambda=0}^{\infty} \sum_{\mu=-\lambda}^{\lambda}
	S_{\ell+\lambda}^{-(m+\mu)} (\bm r)
	R_\lambda^\mu(\varDelta \bm r) 
	.
\end{align}

\subsection{Transformations}
In this section, we list the transformations needed in FMM.
Each translation has a shortened name where the prefix
`P' reads particle or potential, `M' is for multipole moment,
`L' for local expansion, and `2' for `to'.
We assume that multipole moments and local expansions
have finite order $p$, which means $\ell$ takes the range
$0 \le \ell \le p$.

As a special case of (\ref{eq:additionS}),
we have an expansion of $1/r$ potential by
\begin{equation}
\begin{split}
\label{eq:expand}
\frac{1}{\|\bm r_S - \bm r_R\|} 
	&= S_0^{\,0} (\bm r_S - \bm r_R) \\
	&= \sum_{\lambda=0}^{\infty} \sum_{\mu=-\lambda}^{\lambda}
		S_\lambda^{-\mu} (\bm r_S)
		R_\lambda^{\mu} (-\bm r_R)
	\quad\text{for}\quad
	\| \bm r_S \| > \| \bm r_R \|. 
\end{split}
\end{equation}
Thus, if we take the following definition for the multipole moments
\begin{equation}
\label{eq:P2M}
\text{P2M:}\quad
M_\ell^m = \sum_i q_i \cdot R_\ell^{m} (\bm r_M -\bm r_i),
\end{equation}
the potential field outside the multipole sphere
$\Phi(\bm r) = \sum_i (q_i/\|\bm r - \bm r_i\|)$
is given by,
\begin{equation}
\label{eq:M2P}
\text{M2P:}\quad
\Phi(\bm r) = \sum_{\lambda=0}^p \sum_{\mu=-\lambda}^{\lambda}
  M_\lambda^\mu S_\lambda^{-\mu} (\bm r -\bm r_M).
\end{equation}
Here, $\bm r_M$ is the center of expansion,
$\bm r_i$ and $q_i$ the position and charge of particle $i$.
The coefficients of local expansions are available from the
potential field and equation (\ref{eq:extractL}),
which leads directly to a multipole-to-local transformation
\begin{equation}
\label{eq:M2L}
\text{M2L:}\quad
L_\ell^m = \sum_{\lambda = 0}^p \sum_{\mu=-\lambda}^\lambda
  M_\lambda^\mu S_{\ell+\lambda}^{-(m+\mu)} (\bm r_L - \bm r_M).
\end{equation}

From a given set of local expansion coefficients at $\bm r_L$,
the potential field expands to
\begin{equation}
\label{eq:L2P}
\text{L2P:}\quad
\Phi(\bm r) = \sum_{\lambda=0}^p \sum_{\mu=-\lambda}^{\lambda}
  L_\lambda^\mu R_\lambda^\mu (\bm r - \bm r_L),
\end{equation}
and again (\ref{eq:extractL}) makes  a new expansion at $\bm r_{L'}$
(local-to-local transformation) as
\begin{equation}
\label{eq:L2L}
\text{L2L:}\quad
{L'}_\ell^{\, m} = \sum_{\lambda = \ell}^p \sum_{\mu=-\lambda}^\lambda
  L_\lambda^\mu R_{\lambda-\ell}^{\mu-m}(\bm r_{L'} - \bm r_L).
\end{equation}

Finally, we derive a multipole-to-multipole transformation
for a new center $\bm r_{M'}$.
From the  addition theorem (\ref{eq:additionR}) with
$\bm r = \bm r_M - \bm r_i$
and
$\varDelta \bm r = \bm r_{M'} - \bm r_{M}$,
\begin{equation}
R_\ell^m (\bm r_{M'} - \bm r_i) = 
	\sum_{\lambda=0}^{\infty} \sum_{\mu=-\lambda}^{\lambda}
	R_{\ell-\lambda}^{m-\mu} (\bm r_M - \bm r_i)
	R_\lambda^\mu(\bm r_{M'} - \bm r_M) 
	, \nonumber
\end{equation}
and the definition of multipole moments (\ref{eq:P2M}),
the transformation is given by
\begin{equation}
\label{eq:M2M}
\text{M2M:}\quad
{M'}_\ell^{\, m} = \sum_{\lambda = 0}^\ell \sum_{\mu=-\lambda}^\lambda
  M_{\ell-\lambda}^{m-\mu} R_\lambda^{\mu}(\bm r_{M'} - \bm r_M).
\end{equation}
In practice, 
$\mu$ iterates from
$\max(-\lambda, m-(\ell-\lambda))$
to
$\min(\lambda, m+(\ell-\lambda))$.

\subsection{Potential gradient}
In $N$-body simulations, the gradient of the potential is more
important than the potential itself.
We can exploit that the local expansions have the information of
Cartesian gradient of the potential in their first order coefficients.
The expansion is given by
\begin{equation}
\begin{split}
\label{eq:phigrad}
\Phi(\bm r_L + d \bm r)
&=
  L_0^{\,0} 
  -\frac12  \left(L_1^{\,1} - L_1^{-1}\right) dx 
  -\frac i2 \left(L_1^{\,1} + L_1^{-1}\right) dy 
  + L_1^{\,0} \, dz + \ldots \\
&=
  L_0^{\,0} 
  + \left(-\Re L_1^{\,1}\right) dx 
  + \left( \Im L_1^{\,1}\right) dy 
  + L_1^{\,0} \, dz + \ldots
\end{split}
\end{equation}
Here, $\Re$ and $\Im$ are operators to extract the real and 
imaginary part.
Thus, we can use the L2L or M2L implementation instead of L2P or M2P
by setting the destination order as one, when we need the gradient
at the particle position.

\subsection{Computing solid harmonics}
Table \ref{tab:RSlm} motivates us to have a full Cartesian
derivation of the two solid harmonics.
With help of the scaled version of the associated Legendre polynomials
which we define as $\tilde P_\ell^m(r,z) = r^{\ell} \cdot (r \sin\theta)^{-m} \cdot P_\ell^m(\cos\theta)$,
we have
${\tilde P_\ell^m(r,z) \cdot (x+iy)^m} = {r^\ell \cdot P_\ell^m (\cos\theta) \cdot e^{im\phi}}$.
For $m\ \ge 0$, this obeys a recursion,
\begin{equation}
{\tilde P}_{\ell}^m  (r,z) =
\begin{cases}
	(-1)^m (2m-1)!! & (\ell=m) \\ 
	(2\ell-1) z {\tilde P}_{\ell-1}^m (r,z) & (\ell=m+1) \\ 
	\dfrac{2\ell-1} {\ell-m} z   {\tilde P}_{\ell-1}^m (r,z) 
                  -  \dfrac{\ell+m-1}{\ell-m} r^2 {\tilde P}_{\ell-2}^m (r,z) & (\ell \ge m+2)
\end{cases}
. \\
\end{equation}
Now we can compute the solid harmonics by
\begin{align}
R_\ell^m    (x, y, z) &= \frac{1}{(\ell+m)!} Q_\ell^m(x, y, z), \nonumber \\
S_\ell^{-m} (x, y, z) &= \frac{(\ell-m)!}{r^{2\ell+1}} Q_\ell^m(-x, y, -z), \nonumber \\
\text{with}\quad
Q_\ell^m (x, y, z) &= \tilde P_\ell^m(r,z) \cdot (x+iy)^m.
\end{align}
Here, we use octant symmetries,
\begin{equation}
Q_\ell^m (-x,y,-z) = (-1)^\ell Q_\ell^m(x, -y, z) = (-1)^{\ell+m} Q_\ell^{-m} (x, y, z).
\end{equation}

\subsection{Storage format}
All $R_\ell^m$, $S_\ell^m$, $M_\ell^m$, and $L_\ell^m$ 
satisfy the conjugate relation $R_\ell^{-m} = (-1)^m [R_\ell^m]^*$, etc.
Thus, they effectively  have $(p+1)^2$ real numbers when the order
of expansion is $p$.
We save the elements with non-negative $m$ to
one-dimensional arrays $X_i$ with an index
$i = \ell(\ell+1)+m$,
for ranges $0 \le \ell \le p$, $-\ell \le m \le \ell$, and thus 
$0 \le i < (p+1)^2$, as
\begin{equation}
X_{\ell(\ell+1)+m} = 
\begin{cases}
\Re R_\ell^{|m|} & (m \ge 0)\\
\Im R_\ell^{|m|} & (m < 0)
\end{cases}
.
\end{equation}
Note that $S_\ell^{-m}$ for the {M2L} translation in an order $p$ 
calculation requires a size $(2p+1)^2$.

\subsection{Real Matrix form}
Equation (\ref{eq:M2L}) can be regarded as a
linear transformation
$\mathbb R^{(p+1)^2} \rightarrow \mathbb R^{(p+1)^2}$.
First we consider a transformation in complex numbers,
\begin{equation}
\begin{split}
L_\ell^m 
 &= \sum_{\lambda=0}^{p} \sum_{\mu = -\lambda}^{\lambda}
     G_{\ell,\lambda}^{m,\mu} M_\lambda^\mu \\
 &= \sum_{\lambda=0}^{p} \left[
       G_{\ell,\lambda}^{m,0} M_\lambda^0
	   + \sum_{\mu=1}^\lambda 
         \left(G_{\ell,\lambda}^{m,\mu} M_\lambda^\mu + G_{\ell,\lambda}^{m,-\mu} M_\lambda^{-\mu}\right)
     \right],
\end{split}
\end{equation}
with a matrix element $G_{\ell,\lambda}^{m,\mu} \in \mathbb C$
that depends on the relative position of two cells.
From the conjugate relation
$\Re M_\ell^{-m} = (-1)^m \Re M_\ell^m$ 
and
$\Im M_\ell^{-m} = (-1)^{m+1} \allowbreak \Im M_\ell^m$, 
a transformation in real numbers is expressed by
\begin{align*}
\Re L_\ell^m &= \sum_{\lambda=0}^{p} \left[
   A_{\ell,\lambda}^{m,0} M_\lambda^0
	   + \sum_{\mu=1}^\lambda 
	   \left\{ 
	     \left( A_{\ell,\lambda}^{m,\mu} + C_{\ell,\lambda}^{m,\mu}\right)\Re M_\lambda^\mu
	   - \left(B_{\ell,\lambda}^{m,\mu} - D_{\ell,\lambda}^{m,\mu}\right)\Im M_\lambda^\mu 
	   \right\}
 \right],  \\
\Im L_\ell^m &= \sum_{\lambda=0}^{p} \left[
   B_{\ell,\lambda}^{m,0} M_\lambda^0
	   + \sum_{\mu=1}^\lambda 
	   \left\{ 
	     \left(B_{\ell,\lambda}^{m,\mu} + D_{\ell,\lambda}^{m,\mu}\right)\Re M_\lambda^\mu
	   + \left(A_{\ell,\lambda}^{m,\mu} - C_{\ell,\lambda}^{m,\mu}\right)\Im M_\lambda^\mu 
	   \right\}
 \right],  
\end{align*}
with 
\begin{equation}
\begin{aligned}
\label{eq:realmatrix}
A_{\ell,\lambda}^{m,\mu} &= \Re G_{\ell,\lambda}^{m,\mu}, &
B_{\ell,\lambda}^{m,\mu} &= \Im G_{\ell,\lambda}^{m,\mu},\\
C_{\ell,\lambda}^{m,\mu} &= (-1)^\mu \Re G_{\ell,\lambda}^{m,-\mu}, &
D_{\ell,\lambda}^{m,\mu} &= (-1)^\mu \Im G_{\ell,\lambda}^{m,-\mu}.
\end{aligned}
\end{equation}
For $0\le m \le \ell \le p$, this gives a linear transformation
$\mathbb R^{(p+1)^2} \rightarrow \mathbb R^{(p+1)^2}$.

\section{Green's function for periodic boundary conditions}
\label{sec:greenpbc}
In a periodic system, the transformation matrix might be given
by an infinite summation of mirror images, as in
\begin{equation}
\label{eq:infsum}
    G_{\ell,\lambda}^{m,\mu} (\bm r_L - \bm r_M)
	= \sum_{\bm n \in \mathbb Z^3} 
	 S_{\ell+\lambda}^{-(m+\mu)} \bigl(\bm r_L - (\bm r_M + \bm r_{\bm n})\bigr),
\end{equation}
where $\bm r_{\bm n} = (n_x b_x, n_y b_y, n_z b_z)$ is a displacement
of a mirror image for the root box dimension $(b_x, b_y, b_z)$,
and the summation covers  all the three integers $\bm n = (n_x, n_y, n_z) \in \mathbb Z^3$.
The contributions from the  nearest 27 or 125 cells need to be subtracted
which we do not write explicitly under the summation symbol.

\subsection{Infinite summation of periodic FMM}
For the infinite summation of the singular solid harmonics,
we can apply a rapid convergence method for the periodic FMM
\citep{Figueirido:1997:ELS, Amisaki:2000:PEE}.
The basic idea of the method is
based on a splitting with usual and incomplete gamma functions
\[
\gamma(a,x) + \Gamma(a,x) = \Gamma(a), \nonumber
\]
which have definitions for each term as
\begin{equation}
\int_0^x t^{a-1} e^{-t} dt
+
\int_x^\infty t^{a-1} e^{-t} dt
=
\int_0^\infty t^{a-1} e^{-t} dt.
\end{equation}
By substituting $a = \ell+\frac12$ and $x = (\alpha r)^2$, we have a splitting
of a power function
\begin{equation}
\frac{1}{r^{\ell+1}} = 
\frac{\Gamma\left(\ell + \frac12, (\alpha r)^2  \right)}{\Gamma\left(\ell + \frac12 \right)} \frac{1}{r^{\ell+1}} + 
\frac{\gamma\left(\ell + \frac12, (\alpha r)^2  \right)}{\Gamma\left(\ell + \frac12 \right)} \frac{1}{r^{\ell+1}}.
\end{equation}
Then, the latter term with $\gamma()$ is transformed into a
summation in the reciprocal space.
The final form is
\begin{equation}
\begin{split}
\label{eq:amisaki}
\sum_{\bm n \in \mathbb Z^3 \setminus \bm 0}
  S_\ell^m(\bm r_{\bm n})
&=
\sum_{\bm n \in \mathbb Z^3 \setminus \bm 0}
  \frac{\Gamma\left(\ell+\frac12, (\alpha r_{\bm n})^2\right)}
       {\Gamma\left(\ell+\frac12\right)}
  S_\ell^m(\bm r_{\bm n}) \\
&+
\sum_{\bm n \in \mathbb Z^3 \setminus \bm 0}
  \frac{(i\pi)^\ell}{\sqrt\pi}
  \frac{\exp\left(-(\pi k_{\bm n} / \alpha)^2\right)}{\Gamma\left(\ell+\frac12\right)}
  \frac{k_{\bm n}^{2\ell-1}}{V}
  S_\ell^m(\bm k_{\bm n}).
\end{split}
\end{equation}
Here, $\bm k_{\bm n} = (n_x/b_x, n_y/b_y, n_z/b_z)$ is a reciprocal space vector
\footnote{
	We only consider a rectangular box.
},
$V = b_x b_y b_z$ the volume of the root box,
$\alpha$ a splitting parameter,
$r_{\bm n} = \| \bm r_{\bm n} \|$,
and $k_{\bm n} = \| \bm k_{\bm n} \|$.
Both terms decay quickly for increasing $\| \bm n \|$.
Example parameters for a double precision calculation
on a unit box
are $\alpha=1.5$ and $\| \bm n \| \le 4$
for both summations.

\subsection{Green's function for PMMM}
With an offset vector $\bm r$, (\ref{eq:amisaki}) is modified slightly to
\begin{equation}
\begin{split}
\label{eq:greenPBC}
 \sum_{\bm n \in \mathbb Z^3}
  & S_\ell^m(\bm r + \bm r_{\bm n})
= \\
& \sum_{{\bm n \in \mathbb Z^3}}
  \frac{\Gamma\left(\ell+\frac12, \bigl(\alpha \| \bm r + \bm r_{\bm n}\|\bigr)^2\right)}
       {\Gamma\left(\ell+\frac12\right)}
  S_\ell^m(\bm r + \bm r_{\bm n})
+ {} \\
& \sum_{{\bm n \in \mathbb Z^3 \setminus \bm 0}}
  \frac{(-i\pi)^\ell}{\sqrt\pi}
  \frac{\exp\bigl(-(\pi k_{\bm n} / \alpha)^2\bigr)}{\Gamma\left(\ell+\frac12\right)}
  \frac{k_{\bm n}^{2\ell-1}}{V}
  S_\ell^m(\bm k_{\bm n})
  \cdot \exp(2 \pi i \bm k_{\bm n} \cdot \bm r).
\end{split}
\end{equation}
Equality is valid only for $\ell > 2$ and the left-hand side diverges
otherwise.

When $\ell=0$, the right-hand  side of (\ref{eq:greenPBC})  
agrees with the well known Ewald form
\begin{equation}
\begin{split}
&\sum_{\bm n \in \mathbb Z^3}
  \frac{{\rm erfc}\bigl(\alpha \|\bm r + \bm r_{\bm n}\|\bigr)}{\|\bm r + \bm r_{\bm n}\|}
+ {} \\
&\sum_{\bm n \in \mathbb Z^3 \setminus \bm 0}
  \frac{\exp\left(-(\pi k_{\bm n} / \alpha)^2\right)}{\pi V k_{\bm n}^2}
  \cdot \exp(2 \pi i \bm k_{\bm n} \cdot \bm r),
\end{split}
\end{equation}
and $\ell=1$ gives its gradient.
Thus, the right-hand side of (\ref{eq:greenPBC}) could be expected to give an identical
result to the Ewald method. However, it turns out that we need several corrections
in the potential which are due to the conditional convergence at $\ell = 2$.
After several attempts,
the following correction terms are added to the potential:
\begin{equation}\begin{split} \label{eq:phicorr}
\Phi_i^{\text{(corr)}} =& 
  \frac{2\pi}{3V} \sum_{j=1}^{N} q_j \left\| \bm r_i^{\text{(rel)}} - \bm r_j^{\text{(rel)}} \right\|^2 \\
  =& -\frac{4\pi}{3V} \left[ \sum_{j=1}^{N} q_j \bm r_j^{\text{(rel)}} \right] \cdot \bm r_i^{\text{(rel)}} \\
   &+ \frac{2\pi}{3V} \left[ \sum_{j=1}^{N} q_j \left\| \bm r_j^{\text{(rel)}} \right\|^2 \right]
   + \frac{2\pi}{3V} \left[ \sum_{j=1}^{N} q_j \right]  \left\| \bm r_i^{\text{(rel)}} \right\|^2.
\end{split}
\end{equation}
Here, $\bm r_i^{\text{(rel)}}$ is the coordinate of particle $i$
relative to the center of cell in which it resides.
Since the result agrees with the Ewald method when all the particles
reside at the centers of cells, the relative positions 
are involved
in the correction.

An intuitive interpretation of this potential correction is as follows.
The density field which corresponds to the solution of the Ewald method is
\begin{equation} \label{eq:rhoeff}
\rho(\bm r) = \sum_{j=1}^N q_j \left[ \delta \bigl( \bm r - (\bm r_j + \bm r_{\bm n}) \bigr) - \frac{1}{V} \right].
\end{equation}
The last term inside the square bracket is due to the omitted  wave number
$\bm 0$ in the Fourier space.
Now let us examine the potential around $\bm r_j$ 
due to
the uniform counter charge field
$\rho(\bm r) = -q_j/V$.
A total charge in a solid sphere $\{ \bm r \mid \| {\bm r - \bm r_j \| \le \| \bm r_i - \bm r_j \|} \}$
is $-\frac{4\pi}{3V} q_j  \| \bm r_i - \bm r_j \|^3$, and it makes a radial electric field
$-\nabla_i \Phi(\bm r_i) = -\frac{4\pi}{3V} q_j  (\bm r_i - \bm r_j )$ and hence a potential
$\Phi(\bm r_i) = \frac{2\pi}{3V} q_j  \| \bm r_i - \bm r_j \|^2$ which explains (\ref{eq:phicorr}).
Even in a charge neutral system where $\sum_{i=1}^N q_i =0$, the first term of (\ref{eq:phicorr})
still remains
\footnote{
	Consider a two-body charge neutral system with $+q$ at $\bm r_1$ and $-q$ at $\bm r_2$.
	The Ewald method gives a well defined energy that depends on the relative position
	of the two, $U = -q^2 G(\bm r_1 - \bm r_2) = -q^2 G(\bm r_2 - \bm r_1)$ with a Green's
	function $G(\bm r)$, of which the Laplacian is not zero and
	$\nabla^2 G(\bm r) = -4\pi(\delta(\bm r) - 1/V)$.
	Thus, each particle feels the uniform counter charge field of the other,
	even in a charge neutral case of the Ewald method.
}.
Thus, a computation based on a vacuum boundary,
i.e.~$1/r$ potential,
which does not include the contribution from the uniform counter charge field $\rho(\bm r) = -q_j/V$, 
requires the correction (\ref{eq:phicorr}) to obtain an identical potential to the Ewald method
or the solution of the Poisson equation for (\ref{eq:rhoeff}).

See \citet*{deleeuw80a} for a mathematical insight of the correction term.


\subsection{Correction procedure}
The first term of (\ref{eq:phicorr}) can be reflected as a correction
to the first order expansion $L_1^m$ from the summation of the
first order moment $M_1^m$, as in
\begin{equation}\begin{split}
\left[L_1^{\,0}\right]_{\bm i} &:= 
	\left[L_1^{\,0}\right]_{\bm i} + \frac{4 \pi}{3V} \sum_{\bm j} \left[M_1^{\,0}\right]_{\bm j}, \\ 
\left[L_1^{\,1}\right]_{\bm i} &:= 
	\left[L_1^{\,1}\right]_{\bm i} + \frac{8 \pi}{3V} \sum_{\bm j} \left[M_1^{\,1}\right]_{\bm j}^*,
\end{split}\end{equation}
where $\bm i$ or $\bm j$ is an index of a cell, and the summation
on $\bm j$ takes over all the cells.

The second term is a correction to the potential:
\begin{equation}
\left[L_0^{\, 0}\right]_{\bm i} := 
\left[L_0^{\, 0}\right]_{\bm i} -
\frac{2\alpha}{\sqrt\pi}\left[M_0^{\, 0}\right]_{\bm i} +
\frac{2 \pi}{3V} \sum_{\bm j} 
  \left[ \sum_k q_k \left\| \bm r_k^\text{(rel)} \right\|^2 \right]_{\bm j}.
\end{equation}
The self energy term is corrected as well.
The summation $\sum_k q_k \| \bm r_k^\text{(rel)} \|^2$
of each cell has the same dimension as the five quadrupole moments 
$M_2^m\ {(-2 \le m \le 2)}$,
however, is independent from either of the five.

In a charged system where $\sum_{j=1}^N q_j \neq 0$,
the potential of each particle needs an explicit correction together with
a self term:
\begin{equation}\begin{split}
\Phi_i &:= \Phi_i + \frac{2\pi}{3V} \left[ \sum_{j=1}^N q_j \right] 
	\left( \left\| \bm r_i^\text{(rel)} \right\|^2 - \frac{3}{2 \alpha^2} \right), \\
\nabla \Phi_i &:= \nabla \Phi_i + \frac{4\pi}{3V} \left[ \sum_{j=1}^N q_j \right] \bm r_i^\text{(rel)}.
\end{split}\end{equation}

The potential of Ewald summation for the reference value is defined in
\begin{equation}\begin{split}
\label{eq:refewald}
\Phi_i
=&
\sum_{\bm n \in \mathbb Z^3} {\sum_{j=1}^N}
  \frac{{\rm erfc}\bigl(\alpha \|\bm r_{ij} + \bm r_{\bm n}\|\bigr)}{\|\bm r_{ij}+ \bm r_{\bm n}\|}
\\
&+
\sum_{\bm n \in \mathbb Z^3 \setminus \bm 0} 
  \frac{\exp\left(-(\pi k_{\bm n} / \alpha)^2\right)}{\pi V k_{\bm n}^2}
  \sum_{j=1}^N \exp(2 \pi i \bm k_{\bm n} \cdot \bm r_{ij})
\\
&+ \frac{2 \alpha}{\sqrt\pi} q_i -\frac{\pi}{V \alpha^2} \sum_{j=1}^{N} q_j,
\end{split}
\end{equation}
and the total energy $U = \frac12 \sum_{i=1}^N q_i \Phi_i$.
In the first line of (\ref{eq:refewald}), the interaction for $j=i$ is suppressed when $\bm n = \bm 0$.
See also a manual
\footnote{
	\url{http://protomol.sourceforge.net/ewald.pdf}
	[term (7) needs to be doubled)]
}
of {\scshape ProtoMol} framework \citep{Matthey:2004:POF:1024074.1024075}
for the comments on each term.

\section{Array layout of Green's function}
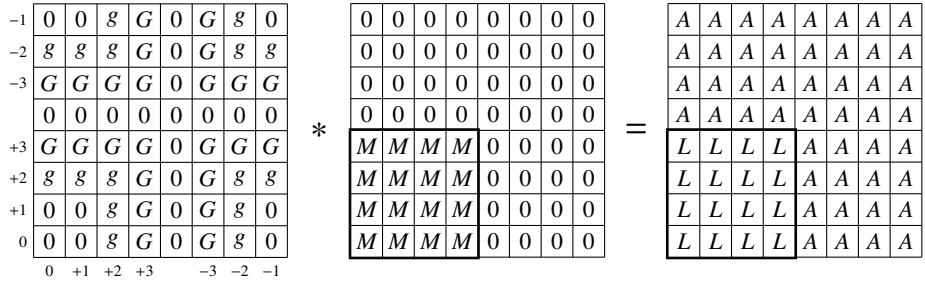
\begin{figure}[htpd]
\centering
\setlength\unitlength{12pt}
\begin{picture}(30,9)(-1,0)

\newsavebox{\boxmm}
\savebox{\boxmm}(8,8)[l]{
  \fontsize{9pt}{9pt}\selectfont{
	\multiput(0,0) (0,1) {9}{\line(1,0){8}}
	\multiput(0,0) (1,0) {9}{\line(0,1){8}}
	\thicklines
	\put(0,0) {\framebox(4,4) {} }

	\multiput(0,0) (1,0) {4} {\makebox(1,1) {$M$} }
	\multiput(0,1) (1,0) {4} {\makebox(1,1) {$M$} }
	\multiput(0,2) (1,0) {4} {\makebox(1,1) {$M$} }
	\multiput(0,3) (1,0) {4} {\makebox(1,1) {$M$} }
	\multiput(0,4) (1,0) {4} {\makebox(1,1) {$0$} }
	\multiput(0,5) (1,0) {4} {\makebox(1,1) {$0$} }
	\multiput(0,6) (1,0) {4} {\makebox(1,1) {$0$} }
	\multiput(0,7) (1,0) {4} {\makebox(1,1) {$0$} }
	\multiput(4,0) (0,1) {8} {\makebox(1,1) {$0$} }
	\multiput(5,0) (0,1) {8} {\makebox(1,1) {$0$} }
	\multiput(6,0) (0,1) {8} {\makebox(1,1) {$0$} }
	\multiput(7,0) (0,1) {8} {\makebox(1,1) {$0$} }
  }
}
\newsavebox{\boxle}
\savebox{\boxle}(8,8)[l]{
  \fontsize{9pt}{9pt}\selectfont{
	\multiput(0,0) (0,1) {9}{\line(1,0){8}}
	\multiput(0,0) (1,0) {9}{\line(0,1){8}}
	\thicklines
	\put(0,0) {\framebox(4,4) {} }

	\multiput(0,0) (1,0) {4} {\makebox(1,1) {$L$} }
	\multiput(0,1) (1,0) {4} {\makebox(1,1) {$L$} }
	\multiput(0,2) (1,0) {4} {\makebox(1,1) {$L$} }
	\multiput(0,3) (1,0) {4} {\makebox(1,1) {$L$} }
	\multiput(0,4) (1,0) {4} {\makebox(1,1) {$A$} }
	\multiput(0,5) (1,0) {4} {\makebox(1,1) {$A$} }
	\multiput(0,6) (1,0) {4} {\makebox(1,1) {$A$} }
	\multiput(0,7) (1,0) {4} {\makebox(1,1) {$A$} }
	\multiput(4,0) (0,1) {8} {\makebox(1,1) {$A$} }
	\multiput(5,0) (0,1) {8} {\makebox(1,1) {$A$} }
	\multiput(6,0) (0,1) {8} {\makebox(1,1) {$A$} }
	\multiput(7,0) (0,1) {8} {\makebox(1,1) {$A$} }
  }
}
\newsavebox{\boxgf}
\savebox{\boxgf}(9,9)[l]{
  \multiput(0,0) (0,1) {9}{\line(1,0){8}}
  \multiput(0,0) (1,0) {9}{\line(0,1){8}}
  \fontsize{6pt}{6pt}\selectfont{
	\put(0,-1) {\makebox(1,0.75)[t]{$0$}}
	\put(1,-1) {\makebox(1,0.75)[t]{$+1$}}
	\put(2,-1) {\makebox(1,0.75)[t]{$+2$}}
	\put(3,-1) {\makebox(1,0.75)[t]{$+3$}}
	\put(5,-1) {\makebox(1,0.75)[t]{$-3$}}
	\put(6,-1) {\makebox(1,0.75)[t]{$-2$}}
	\put(7,-1) {\makebox(1,0.75)[t]{$-1$}}

	\put(-1,0) {\makebox(0.8,1)[r]{$0$}}
	\put(-1,1) {\makebox(0.8,1)[r]{$+1$}}
	\put(-1,2) {\makebox(0.8,1)[r]{$+2$}}
	\put(-1,3) {\makebox(0.8,1)[r]{$+3$}}
	\put(-1,5) {\makebox(0.8,1)[r]{$-3$}}
	\put(-1,6) {\makebox(0.8,1)[r]{$-2$}}
	\put(-1,7) {\makebox(0.8,1)[r]{$-1$}}
  }
  \fontsize{9pt}{9pt}\selectfont{
	\put(0,0) {\makebox(1,1){$0$}}
	\put(1,0) {\makebox(1,1){$0$}}
	\put(2,0) {\makebox(1,1){$g$}}
	\put(3,0) {\makebox(1,1){$G$}}
	\put(4,0) {\makebox(1,1){$0$}}
	\put(5,0) {\makebox(1,1){$G$}}
	\put(6,0) {\makebox(1,1){$g$}}
	\put(7,0) {\makebox(1,1){$0$}}
	\put(0,1) {\makebox(1,1){$0$}}
	\put(1,1) {\makebox(1,1){$0$}}
	\put(2,1) {\makebox(1,1){$g$}}
	\put(3,1) {\makebox(1,1){$G$}}
	\put(4,1) {\makebox(1,1){$0$}}
	\put(5,1) {\makebox(1,1){$G$}}
	\put(6,1) {\makebox(1,1){$g$}}
	\put(7,1) {\makebox(1,1){$0$}}

	\put(0,2) {\makebox(1,1){$g$}}
	\put(1,2) {\makebox(1,1){$g$}}
	\put(2,2) {\makebox(1,1){$g$}}
	\put(3,2) {\makebox(1,1){$G$}}
	\put(4,2) {\makebox(1,1){$0$}}
	\put(5,2) {\makebox(1,1){$G$}}
	\put(6,2) {\makebox(1,1){$g$}}
	\put(7,2) {\makebox(1,1){$g$}}

	\put(0,3) {\makebox(1,1){$G$}}
	\put(1,3) {\makebox(1,1){$G$}}
	\put(2,3) {\makebox(1,1){$G$}}
	\put(3,3) {\makebox(1,1){$G$}}
	\put(4,3) {\makebox(1,1){$0$}}
	\put(5,3) {\makebox(1,1){$G$}}
	\put(6,3) {\makebox(1,1){$G$}}
	\put(7,3) {\makebox(1,1){$G$}}

	\put(0,4) {\makebox(1,1){$0$}}
	\put(1,4) {\makebox(1,1){$0$}}
	\put(2,4) {\makebox(1,1){$0$}}
	\put(3,4) {\makebox(1,1){$0$}}
	\put(4,4) {\makebox(1,1){$0$}}
	\put(5,4) {\makebox(1,1){$0$}}
	\put(6,4) {\makebox(1,1){$0$}}
	\put(7,4) {\makebox(1,1){$0$}}

	\put(0,5) {\makebox(1,1){$G$}}
	\put(1,5) {\makebox(1,1){$G$}}
	\put(2,5) {\makebox(1,1){$G$}}
	\put(3,5) {\makebox(1,1){$G$}}
	\put(4,5) {\makebox(1,1){$0$}}
	\put(5,5) {\makebox(1,1){$G$}}
	\put(6,5) {\makebox(1,1){$G$}}
	\put(7,5) {\makebox(1,1){$G$}}

	\put(0,6) {\makebox(1,1){$g$}}
	\put(1,6) {\makebox(1,1){$g$}}
	\put(2,6) {\makebox(1,1){$g$}}
	\put(3,6) {\makebox(1,1){$G$}}
	\put(4,6) {\makebox(1,1){$0$}}
	\put(5,6) {\makebox(1,1){$G$}}
	\put(6,6) {\makebox(1,1){$g$}}
	\put(7,6) {\makebox(1,1){$g$}}

	\put(0,7) {\makebox(1,1){$0$}}
	\put(1,7) {\makebox(1,1){$0$}}
	\put(2,7) {\makebox(1,1){$g$}}
	\put(3,7) {\makebox(1,1){$G$}}
	\put(4,7) {\makebox(1,1){$0$}}
	\put(5,7) {\makebox(1,1){$G$}}
	\put(6,7) {\makebox(1,1){$g$}}
	\put(7,7) {\makebox(1,1){$0$}}
  }
}

\put( 0,0) {\usebox{\boxgf}}
\put(10,1) {\usebox{\boxmm}}
\put(20,1) {\usebox{\boxle}}
\fontsize{14pt}{14pt}\selectfont{
  \put( 8,4) {\makebox(2,2){$*$}}
  \put(18,4) {\makebox(2,2){$=$}}
}

\end{picture}
\caption{
	A convolution for an open boundary system with $4^2$ cells,
	computing local expansions (right) from a Green's function (left)
	and multipole moments (middle), with $G$ an element of the Green's	
	function, $M$ multipole moments, $L$ local expansions, $A$ an
	aliasing element.
	The asterisk symbol ($*$) denotes a convolution operation.
	The interactions reach up to $\pm 3$ cells for each direction, 
	and the nearest nine interactions are masked with $0$,
	whereas $g$ takes either $G$ or $0$ depending on the cutoff
	distance.
}
\label{fig:green}
\end{figure}

To avoid the aliasing effect,
we need eight times more volume to perform the convolution
operation in a three-dimensional open boundary system.
Figure \ref{fig:green} shows an example layout of the
Green's function in the case of two-dimensional system with
$4^2$ cells.

Here, a convolution operation between two-dimensional arrays is defined by,
\begin{equation}
h = f * g \Leftrightarrow h(i_0, j_0) = \sum_{i,j} f(i_0-i, j_0-j) \times g(i,j)
\end{equation}
with periodic indices, and $f*g = g*f$.

\begin{figure}[htpd]
\centering
\setlength\unitlength{20pt}
\begin{picture}(10,4)(0,0)

\newsavebox{\boxmmrf}
\savebox{\boxmmrf}(4,4)[l]{
  \multiput(0,0) (0,1) {5}{\line(1,0){4}}
  \multiput(0,0) (1,0) {5}{\line(0,1){4}}
  \linethickness{0.1pt}
  \multiput(1,1) (0,0.25) {5}{\line(1,0){1}}
  \multiput(1,1) (0.25,0) {5}{\line(0,1){1}}
  \thicklines
  \put(0,0) {\framebox(3,3) {} }
  \fontsize{14pt}{14pt}\selectfont{
    \put(1,1) {\framebox(1,1) {$\bm M$} }
  }
}

\newsavebox{\boxlerf}
\savebox{\boxlerf}(4,4)[l]{
  \multiput(0,0) (0,1) {5}{\line(1,0){4}}
  \multiput(0,0) (1,0) {5}{\line(0,1){4}}
  \linethickness{0.1pt}
  \multiput(0,0) (0,0.25) {13}{\line(1,0){3}}
  \multiput(0,0) (0.25,0) {13}{\line(0,1){3}}
  \thicklines
  \put(0,0) {\framebox(3,3) {} }
  \multiput(0,1) (0,1) {2}{\line(1,0){3}}
  \multiput(1,0) (1,0) {2}{\line(0,1){3}}
  \fontsize{14pt}{14pt}\selectfont{
	\multiput(0,0) (1,0) {3} {\makebox(1,1) {$\bm L$}}
	\multiput(0,1) (1,0) {3} {\makebox(1,1) {$\bm L$}}
	\multiput(0,2) (1,0) {3} {\makebox(1,1) {$\bm L$}}
  }
  \multiput(0,3) (1,0) {4} {\makebox(1,1) {$A$}}
  \multiput(3,0) (0,1) {3} {\makebox(1,1) {$A$}}
}

\put(0,0) {\usebox{\boxmmrf}}
\put(6,0) {\usebox{\boxlerf}}
\fontsize{14pt}{14pt}\selectfont{
  \put(4,1) {\makebox(2,2) {$\Rightarrow$}}
}

\end{picture}
\caption{
	Computing local expansions of $12^2$ cells ($\bm L$)
	from multipole moments of $4^2$ cells ($\bm M$) , with
	aliasing cells ($A$). Including the margin region, it requires
	$16^2$ transformations.
	The Green's	 function looks quite similar to Fig.~\ref{fig:green}, and
	the interaction reaches	up to $\pm 7$ cells for each direction.
}
\label{fig:refine}
\end{figure}
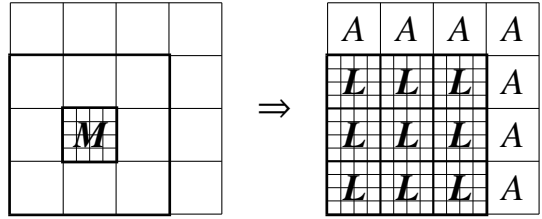

Figure \ref{fig:refine} illustrates a hierarchical treatment
of PMMM in a two-dimensional system.
Contribution of a coarse cell to 9 nearby cells were masked
out. Then, the coarse cell is refined to $4^2$ fine cells,
and contributions of $4^2$ multipole moments to $12^2$ local
expansions are evaluated by a fast convolution method.
Again, contributions of 9 nearby fine cells are masked out.
In this case, $16^2$ FFT and {M2L} transformations are needed
including the margin region. Thus, one fine cell costs the same as 16 transformations. 
This cost does not depend on the size of refined cells.

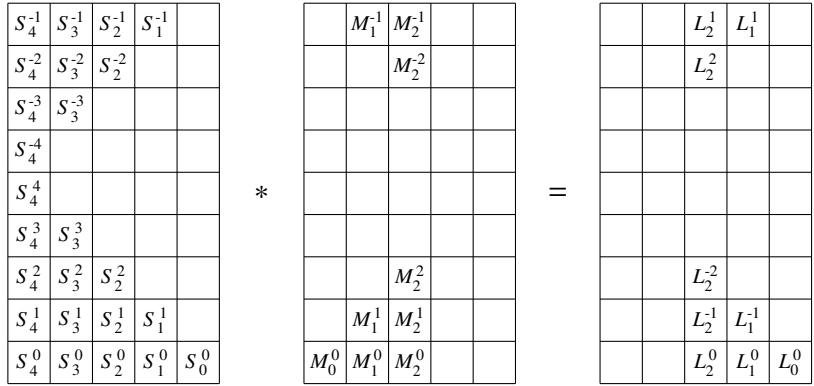
\begin{figure}[htpd]
\centering
\setlength\unitlength{16pt}
\begin{picture}(19,9)(0,0)
\newsavebox{\boxgfmtol}
\savebox{\boxgfmtol}(5,9)[l]{
  \multiput(0,0) (0,1) {10}{\line(1,0){5}}
  \multiput(0,0) (1,0) { 6}{\line(0,1){9}}

  \fontsize{8pt}{8pt}\selectfont{
    \put(4,0) {\makebox(1,1){$S_0^{\,0}$}}

    \put(3,1) {\makebox(1,1){$S_1^{\,1}$}}
    \put(3,0) {\makebox(1,1){$S_1^{\,0}$}}
    \put(3,8) {\makebox(1,1){$S_1^{\mathchar`-1}$}}

    \put(2,2) {\makebox(1,1){$S_2^{\,2}$}}
    \put(2,1) {\makebox(1,1){$S_2^{\,1}$}}
    \put(2,0) {\makebox(1,1){$S_2^{\,0}$}}
    \put(2,8) {\makebox(1,1){$S_2^{\mathchar`-1}$}}
    \put(2,7) {\makebox(1,1){$S_2^{\mathchar`-2}$}}

    \put(1,3) {\makebox(1,1){$S_3^{\,3}$}}
    \put(1,2) {\makebox(1,1){$S_3^{\,2}$}}
    \put(1,1) {\makebox(1,1){$S_3^{\,1}$}}
    \put(1,0) {\makebox(1,1){$S_3^{\,0}$}}
    \put(1,8) {\makebox(1,1){$S_3^{\mathchar`-1}$}}
    \put(1,7) {\makebox(1,1){$S_3^{\mathchar`-2}$}}
    \put(1,6) {\makebox(1,1){$S_3^{\mathchar`-3}$}}

    \put(0,4) {\makebox(1,1){$S_4^{\,4}$}}
    \put(0,3) {\makebox(1,1){$S_4^{\,3}$}}
    \put(0,2) {\makebox(1,1){$S_4^{\,2}$}}
    \put(0,1) {\makebox(1,1){$S_4^{\,1}$}}
    \put(0,0) {\makebox(1,1){$S_4^{\,0}$}}
    \put(0,8) {\makebox(1,1){$S_4^{\mathchar`-1}$}}
    \put(0,7) {\makebox(1,1){$S_4^{\mathchar`-2}$}}
    \put(0,6) {\makebox(1,1){$S_4^{\mathchar`-3}$}}
    \put(0,5) {\makebox(1,1){$S_4^{\mathchar`-4}$}}
  }
}
\newsavebox{\boxmmmtol}
\savebox{\boxmmmtol}(5,9)[l]{
  \multiput(0,0) (0,1) {10}{\line(1,0){5}}
  \multiput(0,0) (1,0) { 6}{\line(0,1){9}}

  \fontsize{8pt}{8pt}\selectfont{
    \put(0,0) {\makebox(1,1){$M_0^{\,0}$}}
    \put(1,8) {\makebox(1,1){$M_1^{\mathchar`-1}$}}
    \put(1,0) {\makebox(1,1){$M_1^{\,0}$}}
    \put(1,1) {\makebox(1,1){$M_1^{\,1}$}}
    \put(2,7) {\makebox(1,1){$M_2^{\mathchar`-2}$}}
    \put(2,8) {\makebox(1,1){$M_2^{\mathchar`-1}$}}
    \put(2,0) {\makebox(1,1){$M_2^{\,0}$}}
    \put(2,1) {\makebox(1,1){$M_2^{\,1}$}}
    \put(2,2) {\makebox(1,1){$M_2^{\,2}$}}
  }
}
\newsavebox{\boxlemtol}
\savebox{\boxlemtol}(5,9)[l]{
  \multiput(0,0) (0,1) {10}{\line(1,0){5}}
  \multiput(0,0) (1,0) { 6}{\line(0,1){9}}

  \fontsize{8pt}{8pt}\selectfont{
    \put(4,0) {\makebox(1,1){$L_0^{\,0}$}}
    \put(3,1) {\makebox(1,1){$L_1^{\mathchar`-1}$}}
    \put(3,0) {\makebox(1,1){$L_1^{\,0}$}}
    \put(3,8) {\makebox(1,1){$L_1^{\,1}$}}
    \put(2,2) {\makebox(1,1){$L_2^{\mathchar`-2}$}}
    \put(2,1) {\makebox(1,1){$L_2^{\mathchar`-1}$}}
    \put(2,0) {\makebox(1,1){$L_2^{\,0}$}}
    \put(2,8) {\makebox(1,1){$L_2^{\,1}$}}
    \put(2,7) {\makebox(1,1){$L_2^{\,2}$}}
  }
}

\put( 0,0) {\usebox{\boxgfmtol}}
\put( 7,0) {\usebox{\boxmmmtol}}
\put(14,0) {\usebox{\boxlemtol}}
  \put( 5,4) {\makebox(2,1){$*$}}
  \put(12,4) {\makebox(2,1){$=$}}

\end{picture}
\caption{
	An {M2L} transformation in a convolution form 
	with an order $p=2$.
}
\label{fig:convmtol}
\end{figure}

The transformation
$L_\ell^m = \sum_{\lambda, \mu} S_{\ell+\lambda}^{-(m+\mu)} M_\lambda^\mu$
itself has also a convolution form.
A possible convolution form
is illustrated in Fig.~\ref{fig:convmtol}.
This reduces the computational complexity of one transformation
from $\mathcal O(p^4)$ to
$\mathcal O(p^2 \log p)$ 
\citep{
	Greengard:1988:EIF,
	Elliott:1996:FFT}.

\section{Use of complex numbers}
A complication arises when both the potential theory
in spherical harmonics and Fourier transformation  are formulated in complex
numbers but what matters is charge and potential in real
numbers.
Several optimizations in this topic have been discussed in the text, 
however, one simple choice is to perform two transformations
simultaneously in full complex operations.
From the linearity, we can exploit
\begin{equation}
L_\ell^{\,m} + i {L'}_\ell^{\,m} = 
\sum_{\lambda,\mu} G_{\ell,\lambda}^{\,m,\mu}
(M_\lambda^{\,\mu} + i {M'}_\lambda^{\,\mu}). 
\end{equation}
First, we compose two sets of coefficients as
\[
C_\ell^m = A_\ell^m + i B_\ell^m,
\]
where
$A_\ell^{-m} = (-1)^m [A_\ell^m]^*$
and
$B_\ell^{-m} = (-1)^m [B_\ell^m]^*$.
The composite $C_\ell^m$ effectively has $2(p+1)^2$ words in real numbers.
Then, from
\begin{equation}\begin{split}
C_\ell^m &= 
    (\Re A_\ell^m - \Im B_\ell^m) + 
   i(\Re B_\ell^m + \Im A_\ell^m ), \\
(-1)^m C_\ell^{-m} &= 
    (\Re A_\ell^m + \Im B_\ell^m) + 
   i(\Re B_\ell^m - \Im A_\ell^m ),
\end{split}\end{equation}
the splitting is given by
\begin{equation}\begin{split}
A_\ell^m &= 
  \frac 1 2 \Re \left[ C_\ell^m + (-1)^m C_\ell^{-m} \right] +
  \frac i 2 \Im \left[ C_\ell^m - (-1)^m C_\ell^{-m} \right],
\\
B_\ell^m &= 
  \frac 1 2 \Im \left[ C_\ell^m + (-1)^m C_\ell^{-m} \right] -
  \frac i 2 \Re \left[ C_\ell^m - (-1)^m C_\ell^{-m} \right].
\end{split}\end{equation}
The equations above can be applied for composing two sets of multipole
moments and splitting the composite of two local expansions.

\bibliographystyle{elsarticle-harv}
\bibliography{fastmultipole,database}

\ifx \undefined \mathbb \def \mathbb #1{{\bf #1}}\fi
\begin{thebibliography}{29}
\expandafter\ifx\csname natexlab\endcsname\relax\def\natexlab#1{#1}\fi
\providecommand{\url}[1]{\texttt{#1}}
\providecommand{\href}[2]{#2}
\providecommand{\path}[1]{#1}
\providecommand{\DOIprefix}{doi:}
\providecommand{\ArXivprefix}{arXiv:}
\providecommand{\URLprefix}{URL: }
\providecommand{\Pubmedprefix}{pmid:}
\providecommand{\doi}[1]{\href{http://dx.doi.org/#1}{\path{#1}}}
\providecommand{\Pubmed}[1]{\href{pmid:#1}{\path{#1}}}
\providecommand{\bibinfo}[2]{#2}
\ifx\xfnm\relax \def\xfnm[#1]{\unskip,\space#1}\fi
\bibitem[{Amisaki(2000)}]{Amisaki:2000:PEE}
\bibinfo{author}{Amisaki, T.}, \bibinfo{year}{2000}.
\newblock \bibinfo{title}{Precise and efficient {Ewald} summation for periodic
  fast multipole method}.
\newblock \bibinfo{journal}{J. Comput. Chem.} \bibinfo{volume}{21},
  \bibinfo{pages}{1075--1087}.
\bibitem[{Bagla(2002)}]{Bagla:2002:TCC}
\bibinfo{author}{Bagla, J.S.}, \bibinfo{year}{2002}.
\newblock \bibinfo{title}{{TreePM}: a code for cosmological {$N$}-body
  simulations}.
\newblock \bibinfo{journal}{Journal of Astrophysics and Astronomy}
  \bibinfo{volume}{23}, \bibinfo{pages}{185--196}.
\bibitem[{Barnes and Hut(1986)}]{Barnes:1986:HFC}
\bibinfo{author}{Barnes, J.E.}, \bibinfo{author}{Hut, P.},
  \bibinfo{year}{1986}.
\newblock \bibinfo{title}{A hierarchical {$O(N\log N)$} force-calculation
  algorithm}.
\newblock \bibinfo{journal}{Nature} \bibinfo{volume}{324},
  \bibinfo{pages}{446--449}.
\bibitem[{Darden et~al.(1993)Darden, York and Pedersen}]{Darden:1993:PME}
\bibinfo{author}{Darden, T.}, \bibinfo{author}{York, D.},
  \bibinfo{author}{Pedersen, L.}, \bibinfo{year}{1993}.
\newblock \bibinfo{title}{Particle mesh {Ewald} --- an {$N \log N$} method for
  {Ewald} sums in large systems}.
\newblock \bibinfo{journal}{J. Chem. Phys.} \bibinfo{volume}{98},
  \bibinfo{pages}{10089--10092}.
\bibitem[{{de Leeuw} et~al.(1980){de Leeuw}, Perram and Smith}]{deleeuw80a}
\bibinfo{author}{{de Leeuw}, S.W.}, \bibinfo{author}{Perram, J.W.},
  \bibinfo{author}{Smith, E.R.}, \bibinfo{year}{1980}.
\newblock \bibinfo{title}{Simulation of electrostatic systems in periodic
  boundary conditions. {I}. lattice sums and dielectric constants}.
\newblock \bibinfo{journal}{Proc. R. Soc. Lond. A} \bibinfo{volume}{373},
  \bibinfo{pages}{27--56}.
\bibitem[{Elliott and {Board}(1996)}]{Elliott:1996:FFT}
\bibinfo{author}{Elliott, W.D.}, \bibinfo{author}{{Board, Jr.}, J.A.},
  \bibinfo{year}{1996}.
\newblock \bibinfo{title}{Fast {Fourier} transform accelerated fast multipole
  algorithm}.
\newblock \bibinfo{journal}{SIAM J. Sci. Comput.} \bibinfo{volume}{17},
  \bibinfo{pages}{398--415}.
\bibitem[{Epton and Dembart(1995)}]{Epton:1995:MTT:210669.210688}
\bibinfo{author}{Epton, M.A.}, \bibinfo{author}{Dembart, B.},
  \bibinfo{year}{1995}.
\newblock \bibinfo{title}{Multipole translation theory for the
  three-dimensional {L}aplace and {H}elmholtz equations}.
\newblock \bibinfo{journal}{SIAM J. Sci. Comput.} \bibinfo{volume}{16},
  \bibinfo{pages}{865--897}.
\bibitem[{Essmann et~al.(1995)Essmann, Perera, Berkowitz, Darden, Lee and
  Pedersen}]{essmann95a}
\bibinfo{author}{Essmann, U.}, \bibinfo{author}{Perera, L.},
  \bibinfo{author}{Berkowitz, M.L.}, \bibinfo{author}{Darden, T.},
  \bibinfo{author}{Lee, H.}, \bibinfo{author}{Pedersen, L.},
  \bibinfo{year}{1995}.
\newblock \bibinfo{title}{A smooth particle mesh {Ewald} method}.
\newblock \bibinfo{journal}{J. Chem. Phys.} \bibinfo{volume}{103},
  \bibinfo{pages}{8577}.
\bibitem[{Figueirido et~al.(1997)Figueirido, Levy, Zholl and
  Berne}]{Figueirido:1997:ELS}
\bibinfo{author}{Figueirido, F.}, \bibinfo{author}{Levy, R.M.},
  \bibinfo{author}{Zholl, R.}, \bibinfo{author}{Berne, B.J.},
  \bibinfo{year}{1997}.
\newblock \bibinfo{title}{Erratum: {``Large scale simulation of macromolecules
  in solution: {Combining} the periodic fast multipole method with multiple
  time step integrators''}}.
\newblock \bibinfo{journal}{J. Chem. Phys.} \bibinfo{volume}{107},
  \bibinfo{pages}{7002--7002}.
\bibitem[{van Gelderen(1998)}]{Gelderen98theshift}
\bibinfo{author}{van Gelderen, M.}, \bibinfo{year}{1998}.
\newblock \bibinfo{title}{The shift operators and translations of spherical
  harmonics}.
\newblock \bibinfo{journal}{DEOS Progress Letter} \bibinfo{volume}{98},
  \bibinfo{pages}{57--67}.
\bibitem[{Greengard(1988)}]{Greengard:1988:REPa}
\bibinfo{author}{Greengard, L.}, \bibinfo{year}{1988}.
\newblock \bibinfo{title}{The rapid evaluation of potential fields in particle
  systems}.
\newblock ACM distinguished dissertations, \bibinfo{publisher}{MIT Press},
  \bibinfo{address}{Cambridge, MA, USA}.
\bibitem[{Greengard and Rokhlin(1987)}]{Greengard:1987:FAP}
\bibinfo{author}{Greengard, L.}, \bibinfo{author}{Rokhlin, V.},
  \bibinfo{year}{1987}.
\newblock \bibinfo{title}{A fast algorithm for particle simulations}.
\newblock \bibinfo{journal}{J. Comput. Phys.} \bibinfo{volume}{73},
  \bibinfo{pages}{325--348}.
\bibitem[{Greengard and Rokhlin(1988a)}]{Greengard:1988:EIF}
\bibinfo{author}{Greengard, L.}, \bibinfo{author}{Rokhlin, V.},
  \bibinfo{year}{1988}a.
\newblock \bibinfo{title}{On the efficient implementation of the fast multipole
  algorithm}.
\newblock \bibinfo{type}{Technical Report} \bibinfo{number}{TR-602}. Yale
  University. \bibinfo{address}{New Haven, CT, USA}.
\bibitem[{Greengard and Rokhlin(1988b)}]{Greengard:1988:REPb}
\bibinfo{author}{Greengard, L.}, \bibinfo{author}{Rokhlin, V.},
  \bibinfo{year}{1988}b.
\newblock \bibinfo{title}{The rapid evaluation of potential fields in three
  dimensions}, in: \bibinfo{editor}{Anderson, C.R.},
  \bibinfo{editor}{Greengard, C.} (Eds.), \bibinfo{booktitle}{{Vortex Methods:
  Proceedings of the U.C.L.A. workshop held in Los Angeles, May 20--22, 1987}},
  \bibinfo{publisher}{Spring{\-}er-Ver{\-}lag}, \bibinfo{address}{Berlin,
  Germany~/ Heidelberg, Germany~/ London, UK~/ etc.}. pp.
  \bibinfo{pages}{121--141}.
\bibitem[{Hesford and Waag(2010)}]{Hesford:2010:FMM}
\bibinfo{author}{Hesford, A.J.}, \bibinfo{author}{Waag, R.C.},
  \bibinfo{year}{2010}.
\newblock \bibinfo{title}{The fast multipole method and {Fourier} convolution
  for the solution of acoustic scattering on regular volumetric grids}.
\newblock \bibinfo{journal}{J. Comput. Phys.} \bibinfo{volume}{229},
  \bibinfo{pages}{8199--8210}.
\bibitem[{Hockney and Eastwood(1988)}]{Hockney:1988:CSU}
\bibinfo{author}{Hockney, R.W.}, \bibinfo{author}{Eastwood, J.W.},
  \bibinfo{year}{1988}.
\newblock \bibinfo{title}{Computer Simulation Using Particles}.
\newblock \bibinfo{publisher}{Adam Hilger Ltd.}, \bibinfo{address}{Bristol,
  UK}.
\bibitem[{Ishiyama et~al.(2009)Ishiyama, Fukushige and
  Makino}]{Ishiyama:2009:GMP}
\bibinfo{author}{Ishiyama, T.}, \bibinfo{author}{Fukushige, T.},
  \bibinfo{author}{Makino, J.}, \bibinfo{year}{2009}.
\newblock \bibinfo{title}{{GreeM}: Massively parallel {TreePM} code for large
  cosmological {$N$}-body simulations}.
\newblock \bibinfo{journal}{Publications of the {Astronomical Society of
  Japan}} \bibinfo{volume}{61}, \bibinfo{pages}{1319--1330}.
\bibitem[{Makino(1999)}]{Makino:1999:YAF}
\bibinfo{author}{Makino, J.}, \bibinfo{year}{1999}.
\newblock \bibinfo{title}{Yet another fast multipole method without
  multipoles---pseudoparticle multipole method}.
\newblock \bibinfo{journal}{J. Comput. Phys.} \bibinfo{volume}{151},
  \bibinfo{pages}{910--920}.
\bibitem[{Makino et~al.(2003)Makino, Fukushige, Koga and
  Namura}]{Makino:2003:GMP}
\bibinfo{author}{Makino, J.}, \bibinfo{author}{Fukushige, T.},
  \bibinfo{author}{Koga, M.}, \bibinfo{author}{Namura, K.},
  \bibinfo{year}{2003}.
\newblock \bibinfo{title}{{GRAPE-6}: Massively-parallel special-purpose
  computer for astrophysical particle simulations}.
\newblock \bibinfo{journal}{Publications of the {Astronomical Society of
  Japan}} \bibinfo{volume}{55}, \bibinfo{pages}{1163--1187}.
\bibitem[{Matthey et~al.(2004)Matthey, Cickovski, Hampton, Ko, Ma, Nyerges,
  Raeder, Slabach and Izaguirre}]{Matthey:2004:POF:1024074.1024075}
\bibinfo{author}{Matthey, T.}, \bibinfo{author}{Cickovski, T.},
  \bibinfo{author}{Hampton, S.}, \bibinfo{author}{Ko, A.}, \bibinfo{author}{Ma,
  Q.}, \bibinfo{author}{Nyerges, M.}, \bibinfo{author}{Raeder, T.},
  \bibinfo{author}{Slabach, T.}, \bibinfo{author}{Izaguirre, J.A.},
  \bibinfo{year}{2004}.
\newblock \bibinfo{title}{Protomol, an object-oriented framework for
  prototyping novel algorithms for molecular dynamics}.
\newblock \bibinfo{journal}{ACM Trans. Math. Softw.} \bibinfo{volume}{30},
  \bibinfo{pages}{237--265}.
\bibitem[{{Message Passing Interface Forum}(2012)}]{mpi-3.0}
\bibinfo{author}{{Message Passing Interface Forum}}, \bibinfo{year}{2012}.
\newblock \bibinfo{title}{{{MPI}: A Message-Passing Interface Standard Version
  3.0}}.
\newblock \bibinfo{note}{Chapter author for Collective Communication, Process
  Topologies, and One Sided Communications}.
\bibitem[{Ong et~al.(2004)Ong, Lee and Lim}]{journals/tcad/OngLL04}
\bibinfo{author}{Ong, E.T.}, \bibinfo{author}{Lee, H.P.}, \bibinfo{author}{Lim,
  K.M.}, \bibinfo{year}{2004}.
\newblock \bibinfo{title}{A parallel fast fourier transform on multipoles
  ({FFTM}) algorithm for electrostatics analysis of three-dimensional
  structures.}
\newblock \bibinfo{journal}{IEEE Trans. Computer-Aided Design}
  \bibinfo{volume}{23}, \bibinfo{pages}{1063--1072}.
\bibitem[{Ong et~al.(2003)Ong, Lim, Lee and Lee}]{Ong:2003:FAT}
\bibinfo{author}{Ong, E.T.}, \bibinfo{author}{Lim, K.M.}, \bibinfo{author}{Lee,
  K.H.}, \bibinfo{author}{Lee, H.P.}, \bibinfo{year}{2003}.
\newblock \bibinfo{title}{A fast algorithm for three-dimensional potential
  fields calculation: fast {Fourier} transform on multipoles}.
\newblock \bibinfo{journal}{J. Comput. Phys.} \bibinfo{volume}{192},
  \bibinfo{pages}{244--261}.
\bibitem[{Sezai et~al.(2007)Sezai, Hisada, Zhai, Chen and
  Sawaya}]{CGFMMFFT2007}
\bibinfo{author}{Sezai, T.}, \bibinfo{author}{Hisada, Y.},
  \bibinfo{author}{Zhai, H.}, \bibinfo{author}{Chen, Q.},
  \bibinfo{author}{Sawaya, K.}, \bibinfo{year}{2007}.
\newblock \bibinfo{title}{Improvement of calculation speed and memory of the
  {MoM} by the {CG-FMM-FFT} method}.
\newblock \bibinfo{journal}{Technical report of IEICE}
  \bibinfo{volume}{SPS2007}, \bibinfo{pages}{7--14}.
\bibitem[{Shan et~al.(2005)Shan, Klepeis, Eastwood, Dror and
  Shaw}]{Shan2005:GSE}
\bibinfo{author}{Shan, Y.}, \bibinfo{author}{Klepeis, J.L.},
  \bibinfo{author}{Eastwood, M.P.}, \bibinfo{author}{Dror, R.O.},
  \bibinfo{author}{Shaw, D.E.}, \bibinfo{year}{2005}.
\newblock \bibinfo{title}{{Gaussian split Ewald: A fast Ewald mesh method for
  molecular simulation}}.
\newblock \bibinfo{journal}{J. Chem. Phys.} \bibinfo{volume}{122},
  \bibinfo{pages}{054101+}.
\bibitem[{Shimada et~al.(1993)Shimada, Kaneko and Takada}]{Shimada:1993:ECC}
\bibinfo{author}{Shimada, J.}, \bibinfo{author}{Kaneko, H.},
  \bibinfo{author}{Takada, T.}, \bibinfo{year}{1993}.
\newblock \bibinfo{title}{Efficient calculations of {Coulombic} interactions in
  biomolecular simulations with periodic boundary conditions}.
\newblock \bibinfo{journal}{J. Comput. Chem.} \bibinfo{volume}{14},
  \bibinfo{pages}{867--878}.
\bibitem[{Shimada et~al.(1994)Shimada, Kaneko and Takada}]{Shimada:1994:PFM}
\bibinfo{author}{Shimada, J.}, \bibinfo{author}{Kaneko, H.},
  \bibinfo{author}{Takada, T.}, \bibinfo{year}{1994}.
\newblock \bibinfo{title}{Performance of fast multipole methods for calculating
  electrostatic interactions in biomacromolecular simulations}.
\newblock \bibinfo{journal}{J. Comput. Chem.} \bibinfo{volume}{15},
  \bibinfo{pages}{28--43}.
\bibitem[{Springel(2005)}]{Springel:2005:CSC}
\bibinfo{author}{Springel, V.}, \bibinfo{year}{2005}.
\newblock \bibinfo{title}{The cosmological simulation code {GADGET-2}}.
\newblock \bibinfo{journal}{Monthly Notices of the Royal Astronomical Society}
  \bibinfo{volume}{364}, \bibinfo{pages}{1105--1134}.
\bibitem[{Wang and LeSar(1996)}]{Wang:1996:EFM}
\bibinfo{author}{Wang, H.Y.}, \bibinfo{author}{LeSar, R.},
  \bibinfo{year}{1996}.
\newblock \bibinfo{title}{An efficient fast multipole algorithm based on an
  expansion in the solid harmonics}.
\newblock \bibinfo{journal}{J. Chem. Phys.} \bibinfo{volume}{104},
  \bibinfo{pages}{4173--4179}.

\end{thebibliography}







\end{document}